\documentclass[11pt]{article}
\usepackage{amssymb}

\usepackage{amsmath}

\newcounter{resultnum}[section]

\newcounter{conclusionnum}[section]

\newcounter{conditionnum}[section]

\newcounter{conjecturenum}[section]
\newtheorem{example}{Example}[section]

\newcounter{examplenum}[section]

\newcounter{exercisenum}[section]
\newtheorem{lemma}{Lemma}[section]

\newcounter{lemmanum}[section]

\newcounter{notationnum}[section]
\newtheorem{theorem}{Theorem}[section]

\newcounter{theoremnum}[section]
\newtheorem{definition}{Definition}[section]

\newcounter{definitionnum}[section]
\newtheorem{corollary}{Corollary}[section]

\newcounter{corollarynum}[section]

\newcounter{remarknum}[section]
\newtheorem{proposition}{Proposition}[section]

\newcounter{propositionnum}[section]

\newcounter{acknowledgementnum}[section]

\newcounter{algorithmnum}[section]

\newcounter{axiomnum}[section]

\newcounter{casenum}[section]

\newcounter{claimnum}[section]

\newcounter{summarynum}[section]

\newcounter{problemnum}[section]
\newenvironment{proof}[1][]{\textbf{Proof.} }{}

\begin{document}

\title{Einstein Gravity in Almost K\"{a}hler and \\
Lagrange--Finsler Variables and \\
Deformation Quantization}
\date{April 5, 2010}
\author{Sergiu I. Vacaru\thanks{
e--mails: sergiu.vacaru@uaic.ro,\ Sergiu.Vacaru@gmail.com,\ \newline
http://www.scribd.com/people/view/1455460-sergiu} \\
{\quad} \\
{\small {\textsl{Science Department, University "Al. I. Cuza" Ia\c si,}} }\\
{\small {\textsl{\ 54 Lascar Catargi street, 700107, Ia\c si, Romania}} }}
\maketitle

\begin{abstract}
A geometric procedure is elaborated for transforming (pseudo) 
Riemanian met\-rics and connections into canonical geometric 
objects (met\-ric and nonlinear and linear connections) for 
effective Lagrange, or Fin\-s\-ler,
geometries which, in turn, can be equivalently represent\-ed 
as almost K\"{a}hler spaces. This allows us to formulate an approach to quantum gravity following standard methods of deformation quantization. Such constructions
are performed not on tangent bundles, as in usual Finsler geometry, but on spacetimes enabled with nonholonomic distributions defining 2+2 splitting with associate nonlinear connection structure. We also show how the Einstein
equations can be written in terms of Lagrange--Finsler variables and corresponding almost symplectic structures and encoded into the zero--degree cohomology coefficient for a quantum model of Einstein manifolds.

\vskip3pt

\textbf{Keywords:}\ Einstein spaces, Lagrange geometry, Finsler geometry,
deformation quantization, quantum gravity

\vskip3pt

MSC:\ 83C99, 53D55, 53B40, 53B35

PACS:\ 04.20.-q, 02.40.-k, 02.90.+g, 02.40.Yy
\end{abstract}


\newpage

\section{Introduction}

The ideas presented in this article grew out of our attempts \cite%
{vqgr1,vqgr2,vqgr3,vbrane} to reformulate equivalently the Einstein gravity
and generalizations as almost K\"{a}hler geometries in order to apply the
deformation quantization methods \cite{fedos1,fedos2,konts1,konts2} and
elaborate quantum models of such spaces. The term ''quantization'' usually
means a procedure stating a quantum model for a given classical theory. One
follows certain analogy with existing approaches to quantum and classical
mechanics and field theories. Such constructions involve a great amount of
ambiguity because quantum theories provide more refined descriptions of
physical systems than the classical ones. For gravity and gauge theories,
this is related to a set of unsolved yet fundamental mathematical problems
in nonlinear functional analysis and noncommutative geometry.

A deeper level of description of quantum systems connected to a consistent
quantization formalism manifests itself in extra geometric structures to be
defined on a generalized phase space. For instance, a symplectic connection
and related Poisson structures, which are not restricted to classical
dynamics, become key ingredients of geometric quantum theories. The existing
examples of deformation quantization of gravity and matter fields involve
more ''rigid'' geometric structures (like metric and nonlinear and linear
connections, torsions, almost complex structures etc). Different choices for
such structures lead, in general, to inequivalent quantizations on the same
phase space, a typical feature of nonlinear theories.

This work was stimulated in part by many attempts to apply deformation
quantization to generalized Poisson structures, gravity and string theory.
Here we mention some approaches and emphasize the following key ideas and
results:

A formal scheme proposing a unification of the four dimensional Einstein
gravity and quantum mechanics was proposed, which resulted in loop quantum
gravity (LQG) and spin network theory, see comprehensive reviews in \cite%
{rovelli,asht3,thiem1}. We also cite a discussion \cite{smolin2} on
alternative approaches following methods which differ from those proposed in
string theory \cite{string1,string2,string3}, which aim to formulate a
unified higher dimensional theory of interactions. The background free LQG
originated from a canonical formulation of general relativity, based on ADM
3+1 decomposition and the Palatini formalism (see a review of results in %
\cite{mtw}), and Ashtekar's connection dynamics with a relatively simple
Hamiltonian constraint and further modifications.

However, the attempts to quantize nonlinear field theories, including
different models of gravity, were oriented not only to the canonical quantum
gravity and/or LQG but also to alternative methods of geometric
quantization. The fundamental works \cite{berez,bffls1,bffls2} preceding the
Fedosov and Kontsevich approaches \cite{fedos1,fedos2,konts1,konts2} should
also be mentioned. In Ref. \cite{gcpp1}, a deformation quantization for
self--dual gravity formulated in Pleba\'{n}ski (self--dual) variables was
considered. A variant of deformation (Moyal) quantization for general
constrained Hamiltonian systems was introduced in \cite{antonsen1}. It was
shown in \cite{antonsen1} how second class constraints can be turned into
first class quantum constraints. The conditions pointing to existence of
anomalies were derived and it was analyzed how some kinds of anomalies can
be removed. The conclusion was that the deformation quantization of pure
Yang--Mills theory is straightforward whereas gravity is anomalous. It was
also stated that in the ADM formalism of gravity, the anomaly is very
complicated and the equations picking out physical states become infinite
order functional differential equations. The Ashtekar variables remedy both
of these problems in 2+1 dimensions but not in 3+1 dimensions. Recently, the
deformation quantization was applied to linearized Einstein's equations \cite%
{qut} using the analogy with Maxwell equations.

There is a series of works by C. Castro oriented to deformation quantization
of gravity for strings and membranes contained in higher dimensions (in a
sense, it is a realization of holographic idea): In Ref. \cite{castro1}, a
geometric derivation of $W_{\infty }$ gravity based on Fedosov's deformation
quantization of symplectic manifolds was elaborated. A holographic reduction
of higher dimensional gravity was attained \cite{castro2} based on the
result of Ref. \cite{cho} that $m+n$ dimensional Einstein gravity can be
identified with an $m$--dimensional generally invariant gauge theory of $%
Diffs N$ (where $N$ is an $n$--dimensional internal manifold), which allowed
a deformation of gravity via Moyal noncommutative star products associated
with the lower dimensional $SU (\infty) $ gauge theory. In \cite{castro3} it
was argued how a quantization of four dimensional gravity could be attained
via a two dimensional quantum $W_\infty$ gauge theory coupled to an
infinite--component sacalar--multiplet. It was shown also how strings and
membranes actions in two and three dimensions emerge from four dimensional
Einstein gravity by using the nonlinear connection formalism of
Lagrange--Finsler and Hamilton--Cartan spaces (the formalism was also
recently considered in quantum gravity in Refs. \cite{vqgr1,vqgr2,vqgr3},
see also the approaches with nonlinear connections to geometric mechanics %
\cite{ma1987,ma,vbrane} and applications in modern commutative and
noncommutative gravity and string theories \cite{vsgg,vrfg}).

The present work belongs to a series of papers on Fedosov quantization of
nonholonomic manifolds and generalized Lagrange--Finsler and Einstein spaces %
\cite{vqgr1,vqgr2,vqgr3,vbrane,avdq2,esv}. Our main idea is to use such
nonholonomic spacetime decompositions (for four dimensions, stating 2+2
splitting), with associated nonlinear connection (N--connection) induced by
certain off--diagonal metric coefficients, when almost symplectic structures
can be defined naturally for (pseudo) Riemannian spaces. Instead of the
Levi--Civita connection, it is convenient to work with another class of
metric--compatible linear connections adapted to the N--connection
structure. Such connections are also uniquely defined by the metric
structure but contain some nontrivial torsion components induced effectively
by generic off--diagonal metric coefficients. The formalism allows us to
quantize similarly both the Lagrange--Finsler and Einstein gravity. The
final constructions can be redefined in terms of the Levi--Civita connection.

In Ref. \cite{vqgr2}, we proved that the Einstein gravity can be lifted
canonically to the tangent bundle and transformed into an almost K\"{a}hler
structure which can be quantized, for instance, following a generalization
of Fedosov's method by Karabegov and Schlichenmaier \cite{karabeg1}. It was
emphasized that the models of quantum gravity on tangent bundles (or
effectively constructed on higher dimension spacetimes) result in violation
of local Lorentz symmetry. In addition to tangent bundles (extra dimensions)
quantum deformation approaches to gravity, models with violated/restricted
fundamental symmetries also present substantial interest in modern physics.
In Ref. \cite{vqgr3}, we argued that Einstein's gravity can be quantized by
nonholonomic deformations with effective generalized Lagrangians. This was
done by transforming (semi) Riemannian metrics and connections into
equivalent (almost symplectic) forms and connections adapted to a canonical
N--connection structure on the same manifold, when the local Lorentz
symmetry is preserved both for the classical and quantized models; see Ref. %
\cite{vqgr3} for further details.

The aim of this work is twofold: First, to show how by nonholonomic frame
transforms any (pseudo) Riemannian metric and corresponding Levi--Civita
connection can be transformed into a canonical metric and nonlinear and
linear connection structures for a Lagrange or Finsler geometry. We will
also develop a technique for constructing an effective regular Lagrangian
generated from a general metric structure with a formal 2+2 spacetime
decomposition. Second, to prove that such effective Lagrange--Finsler spaces
can be equivalently written in terms of an almost K\"{a}hler geometry. It is
also analyzed how quantum properties of gravitational fields and motion
equations are encoded into the 2--forms related the Chern--Weyl
cohomological forms.

The work is organized as follows:

In section 2, we prove that any (pseu\-do) Ri\-e\-man\-nian metric can be
represented in a form similar to canonical metrics in Lagrange or Finsler
geometry by the corresponding frame transform. We provide an analogous
re--formulation of metric and frame variables from Einstein's gravity in
Lagrange--Finsler type variables with an effective Lagrange or Finsler
generating function defining a canonical 2+2 splitting of four dimensional
spacetimes.

Section 3 is devoted to almost K\"{a}hler models of (pseudo) Riemannian and
Lagrange spaces. Canonical almost symplectic forms and connections can be
generated both by the metric structure and a nonholonomic distribution
introduced on the original spacetime. We discuss the similarity and
differences of fundamental geometric objects for the approaches with $3+1 $
fibrations and $2+2$ nonholonomic splittings in gravity theories.

The Fedosov's operators are generalized in section 4 for nonholonomic
manifolds and almost K\"{a}hler structures defined by effective Lagrange, or
Finsler, fundamental functions on (pseudo) Riemannian manifolds.

In section 5, the main results on deformation quantization of almost K\"{a}%
hler geometries are reformulated for (pseudo) Riemannian nonholonomic
manifolds and related Lagrange--Finsler spaces. We show how the information
about Einstein's equations is encoded into the zero--degree cohomology
coefficient of the correspondingly quantized Einstein's manifolds.

Finally, in section 6, we conclude and discuss the results.

\section{(Pseudo)Riemannian Spaces Modeling La\-gran\-ge and Fins\-ler
Geo\-metries}

Let us consider a real (pseudo) Riemann manifold $V^{2n}$ of necessary
smooth class; $\dim V^{2n}=2n,$ where the dimension $n\geq 2$ is fixed.%
\footnote{%
for constructions related to Einstein's gravity $2n=4$} We label the local
coordinates in the form $u^{\alpha }=(x^{i},y^{a}),$ or $u=(x,y),$ where
indices run values $i,j,...=1,2,...n$ and $a,b,...=n+1,n+2,...,n+n,$ and $%
x^{i}$ and $y^{a}$ are respectively the conventional horizontal / holonomic
(h) and vertical / nonholonomic coordinates (v). For the local Euclidean
signature, we consider that all local basis vectors are real but, for the
pseudo--Euclidean signature $(-,+,+,+),$ we introduce $e_{j=1}=\mathit{i}%
\partial /\partial x^{1},$ where $i$ is the imaginary unity, $\mathit{i}%
^{2}=-1,$ and the local coordinate basis vectors can be written in the form $%
e_{\alpha }=\partial /\partial u^{\alpha }=(\mathit{i}\partial /\partial
x^{1},\partial /\partial x^{2},...,\partial /\partial x^{n},\partial
/\partial y^{a}).$\footnote{%
for simplicity, we shall omit to write in explicit form the imaginary unity
considering that we can always distinguish the pseudo--Euclidean signature
by a corresponding metric form or a local system of coordinates with a
coordinate proportional to the imaginary unit} The Einstein's rule on
summing up/low indices will be applied unless indicated otherwise.

Any metric on $V^{2n}$ can be written as%
\begin{equation}
\mathbf{g}=g_{ij}(x,y)\ e^{i}\otimes e^{j}+h_{ab}(x,y)\ e^{a}\otimes \ e^{b},
\label{m1}
\end{equation}%
where the dual vielbeins (tetrads, in four dimensions) $e^{a}=(e^{i},e^{a})$
are parametrized as
\begin{equation}
e^{i}=e_{\ \underline{i}}^{i}(u)dx^{\underline{i}}\mbox{ and }e^{a}=e_{\
\underline{i}}^{a}(u)dx^{\underline{i}}+e_{\ \underline{a}}^{a}(u)dy^{%
\underline{a}},  \label{m1fr}
\end{equation}%
for $e_{\underline{\alpha }}=\partial /\partial u^{\underline{\alpha }}=(e_{%
\underline{i}}=\partial /\partial x^{\underline{i}},e_{\underline{a}%
}=\partial /\partial y^{\underline{a}})$ and $e^{\underline{\beta }}=du^{%
\underline{\beta }}=(e^{\underline{j}}=dx^{\underline{j}},dy^{\underline{b}%
}) $ being, respectively, any fixed local coordinate base and dual base. One
can also consider coordinate transforms such as $u^{\alpha ^{\prime
}}=u^{\alpha ^{\prime }}(u^{\alpha })=\left( x^{i^{\prime }}(u^{\alpha
}),y^{a^{\prime }}(u^{\alpha })\right) .$

\begin{proposition}
\label{pr01}Any metric $\mathbf{g}$ (\ref{m1}) can be expressed in the form
\begin{equation}
\mathbf{\check{g}}=\check{g}_{i^{\prime }j^{\prime }}(x,y)\ \check{e}%
^{i^{\prime }}\otimes \check{e}^{j^{\prime }}+\check{h}_{a^{\prime
}b^{\prime }}(x,y)\ \mathbf{\check{e}}^{a^{\prime }}\otimes \ \mathbf{%
\check{e}}^{b^{\prime }},  \label{hvmetr1}
\end{equation}%
where $\check{e}^{i^{\prime }}=\delta _{\underline{i}}^{i^{\prime }}dx^{%
\underline{i}}$ and $\mathbf{\check{e}}^{a^{\prime }}=\delta _{\ \underline{a%
}}^{a^{\prime }}(u)dy^{\underline{a}}+\check{N}_{\ \underline{i}}^{a^{\prime
}}(u)dx^{\underline{i}}$ for
\begin{eqnarray}
\check{h}_{a^{\prime }b^{\prime }}(u) &=&\frac{1}{2}\frac{\partial ^{2}%
\mathcal{L}(x^{i^{\prime }},y^{c^{\prime }})}{\partial y^{a^{\prime
}}\partial y^{b^{\prime }}},  \label{elf} \\
\check{N}_{\ \underline{i}}^{a^{\prime }}(u) &=&\frac{\partial G^{a^{\prime
}}(x,y)}{\partial y^{n+\underline{j}}},  \label{ncel}
\end{eqnarray}%
where $\delta _{\underline{i}}^{i^{\prime }}$ is the Kronecker symbol, $%
\check{g}_{i^{\prime }j^{\prime }}=\check{h}_{n+i^{\prime }\ n+j^{\prime }}$
and $\check{h}^{ab}$ is the inverse of $\check{h}_{a^{\prime }b^{\prime }},$
for $\det |\check{h}_{a^{\prime }b^{\prime }}|\neq 0$ and
\begin{equation}
2G^{a^{\prime }}(x,y)=\frac{1}{2}\ \check{h}^{a^{\prime }\ n+i}\left( \frac{%
\partial ^{2}\mathcal{L}}{\partial y^{i}\partial x^{k}}y^{n+k}-\frac{%
\partial \mathcal{L}}{\partial x^{i}}\right) ,  \label{sprlf}
\end{equation}%
where $i,k=1,2,...n.$
\end{proposition}

\begin{proof}
Let us fix a generating real function $\mathcal{L}(x^{i^{\prime
}},y^{c^{\prime }}),$ called effective Lagrangian, and compute the values (%
\ref{elf}), (\ref{sprlf}) and (\ref{ncel}), i.e., let's define the
coefficients of metric (\ref{hvmetr1}) with respect to the local coordinate
basis $du^{\underline{\alpha }}=(dx^{\underline{i}},dy^{\underline{a}}).$
Then we introduce the vielbein coefficients (\ref{m1fr}) in (\ref{m1}) and
regroup the coefficients with respect to $du^{\underline{\alpha }}=(dx^{%
\underline{i}},dy^{\underline{a}}).$ Both formulas (\ref{m1}) and (\ref%
{hvmetr1}) define the same metric structure, i.e., $\mathbf{g}=\mathbf{%
\check{g},}$ if the following conditions are satisfied
\begin{equation}
g_{ij}e_{\ \underline{i}}^{i}e_{\ \underline{j}}^{j}=\check{g}_{\underline{i}%
\underline{j}},\ h_{ab}e_{\ \underline{a}}^{a}e_{\ \underline{b}}^{b}=\check{%
h}_{a^{\prime }b^{\prime }}\delta _{\ \underline{a}}^{a^{\prime }}\delta _{\
\underline{b}}^{b^{\prime }},\ h_{ab}e_{\ \underline{i}}^{a}e_{\ \underline{j%
}}^{b}=\check{h}_{a^{\prime }b^{\prime }}\check{N}_{\ \underline{i}%
}^{a^{\prime }}\check{N}_{\ \underline{j}}^{b^{\prime }}.  \label{aleq}
\end{equation}%
In four dimensions, $n=2,$ we have an algebraic system of 6 equations (\ref%
{aleq}) for 12 unknown variables $e_{\ \underline{i}}^{i},e_{\ \underline{i}%
}^{a}$ and $e_{\ \underline{a}}^{a},$ for $\check{g}_{i^{\prime }j^{\prime
}},\check{h}_{a^{\prime }b^{\prime }}$ and $\check{N}_{\ \underline{a}%
}^{a^{\prime }}$ prescribed by $\mathcal{L}$. Such a system can be solved in
order to find the 10 independent coefficients of any $\mathbf{g=\{}g_{\alpha
\beta}\}$ (if this metric is a solution of the Einstein equations, there are
only 6 independent coefficients, because of the Bianchi identities; four of
the metric coefficients can be transformed to zero by use of the
corresponding coordinate transforms). To build inverse constructions, we can
prescribe the coefficients $g_{\alpha \beta }$ by taking any effective
generating function $\mathcal{L}(x^{i^{\prime }},y^{c^{\prime }})$ when the
system (\ref{aleq}) has nontrivial solutions (for simplicity, we can
consider only real vielbeins and local coordinate bases adapted to the
spacetime signature). $\square $
\end{proof}

\vskip5pt

By a straightforward computation one proves:

\begin{lemma}
\label{lem01}Considering $\mathcal{L}$ from (\ref{elf}) and (\ref{sprlf}) to
be a regular Lagrangian, we have that the Euler--Lagrange equations
\begin{equation}
\frac{d}{d\tau }\left( \frac{\partial \mathcal{L}}{\partial y^{i}}\right) -%
\frac{\partial \mathcal{L}}{\partial x^{i}}=0,  \label{eleq}
\end{equation}%
where $y^{i}=y^{n+i}=\frac{dx^{i}}{d\tau }$ for $x^{i}(\tau )$ depending on
the parameter $\tau$. The above Euler--Lagrange equations are equivalent to
the ``nonlinear'' geodesic equations
\begin{equation}
\frac{d^{2}x^{i}}{d\tau ^{2}}+2G^{i}(x^{k},\frac{dx^{j}}{d\tau })=0
\label{ngeq}
\end{equation}%
defining the paths of a canonical semispray $S=y^{i}\frac{\partial }{%
\partial x^{i}}-2G^{a}(x,y)\frac{\partial }{\partial y^{a}},$ for $G^{a}$
given by equations (\ref{sprlf}).
\end{lemma}

The Lemma motivates:

\begin{definition}
A (pseudo) Riemannian space with metric $\mathbf{g}$ (\ref{m1}) is modelled
by a mechanical system with regular effective Lagrangian $\mathcal{L}$ if
there is a nontrivial frame transform defined by any $e_{\ \underline{i}%
}^{i},e_{\ \underline{i}}^{a}$ and $e_{\ \underline{a}}^{a}$ when $\mathbf{g}%
=\mathbf{\check{g}}$\ (\ref{hvmetr1}).
\end{definition}

Inversely, we say that a regular mechanical model with Lagrangian $\mathcal{L%
}$ and Euler--Lagrange equations (\ref{eleq}) is geometrized in terms of a
(pseudo) Riemannian geometry with metric $\mathbf{g}$ (\ref{m1}) if $%
\mathcal{L}$ is a generating function for (\ref{elf}), (\ref{sprlf}) and (%
\ref{ncel}), when $\mathbf{g}=\mathbf{\check{g}}$ (\ref{hvmetr1}) and the
nonlinear geodesic equations (\ref{ngeq}) are equivalent to (\ref{eleq}).

Any equivalent modelling of regular mechanical systems as (pseudo)
Riemannian spaces introduces additional geometric structures on a manifold $%
V^{2n}.$

\begin{definition}
\label{defnc}A nonlinear connection (N--connection) $\mathbf{N}$ on $V^{2n}$
is defined by a Whitney sum (nonholonomic distribution)
\begin{equation}
T(V^{2n})=h(V^{2n})\oplus v(V^{2n}),  \label{whitney}
\end{equation}%
splitting globally the tangent bundle $T(V^{2n})$ into respective h-- and
v--subspac\-es, $h(V^{2n})$ and $v(V^{2n}),$ given locally by a set of
coefficients $N_{i}^{a}(x,y)$ where
\begin{equation*}
\mathbf{N=}N_{i}^{a}(x,y)dx^{i}\otimes \frac{\partial }{\partial y^{a}}.
\end{equation*}
\end{definition}

We note that a subclass of linear connections is defined by $%
N_{i}^{a}=\Gamma _{b}^{a}(x)y^{b}.$

Having prescribed on a $\mathbf{V}^{2n}$ a N--connection structure $\mathbf{%
N=\{}N_{j}^{a}\mathbf{\},}$ we can define a preferred frame structure (with
coefficients depending linearly on $N_{j}^{a})$ denoted $\mathbf{e}_{\nu }=(%
\mathbf{e}_{i},e_{a}),$ where
\begin{equation}
\mathbf{e}_{i}=\frac{\partial }{\partial x^{i}}-N_{i}^{a}(u)\frac{\partial }{%
\partial y^{a}}\mbox{ and
}e_{a}=\frac{\partial }{\partial y^{a}},  \label{dder}
\end{equation}%
with dual frame (coframe) structure $\mathbf{e}^{\mu }=(e^{i},\mathbf{e}%
^{a}),$ where
\begin{equation}
e^{i}=dx^{i}\mbox{ and }\mathbf{e}^{a}=dy^{a}+N_{i}^{a}(u)dx^{i},
\label{ddif}
\end{equation}%
satisfying nontrivial nonholonomy relations
\begin{equation}
\lbrack \mathbf{e}_{\alpha },\mathbf{e}_{\beta }]=\mathbf{e}_{\alpha }%
\mathbf{e}_{\beta }-\mathbf{e}_{\beta }\mathbf{e}_{\alpha }=W_{\alpha \beta
}^{\gamma }\mathbf{e}_{\gamma }  \label{anhrel}
\end{equation}%
with (antisymmetric) anholonomy coefficients $W_{ia}^{b}=\partial
_{a}N_{i}^{b}$ and $W_{ji}^{a}=\Omega _{ij}^{a}.$

Here boldface symbols are used for the spaces with N--connection structure
and for the geometric objects adapted (N--adapted) to the h-- and
v--splitting (\ref{whitney}). We can perform N--adapted geometric
constructions by defining the coefficients of geometric objects (and
associated equations) with respect to noholonomic frames of type (\ref{dder}%
) and (\ref{ddif}). The N--adapted tensors, vectors, forms, etc., are called
respectively distinguished tensors, etc., (in brief, d--tensors, d--vectors,
d--forms, etc.). For instance, a vector field $\mathbf{X}\in T\mathbf{V}^{2n}
$ is expressed as $\mathbf{X}=(hX,\ vX),$ or $\mathbf{X}=X^{\alpha }\mathbf{e%
}_{\alpha }=X^{i}\mathbf{e}_{i}+X^{a}e_{a},$ where $hX=X^{i}\mathbf{e}_{i}$
and $vX=X^{a}e_{a}$ state, respectively, the horizontal (h) and vertical (v)
components of the vector adapted to the N--connection structure.

\begin{proposition}
\label{pr02}Any effective regular Lagrangian $\mathcal{L}$, prescribed on $%
\mathbf{V}^{2n}$, defines a canonical N--connection structure $\mathbf{%
\check{N}=\{}\check{N}_{\ \underline{i}}^{a^{\prime }}(u)\}$ (\ref{ncel})
and preferred frame structures $\mathbf{\check{e}}_{\nu }=(\mathbf{\check{e}}%
_{i},e_{a^{\prime }})$ and $\mathbf{\check{e}}^{\mu }=(e^{i},\mathbf{%
\check{e}}^{a^{\prime }}).$
\end{proposition}

\begin{proof}
The proposition can be proved by straightforward computations. The
coefficients $\check{N}_{\ \underline{i}}^{a^{\prime }}$ satisfy the
conditions of Definition \ref{defnc}. We define $\mathbf{\check{e}}_{\nu }=(%
\mathbf{\check{e}}_{i},e_{a})$ and $\mathbf{\check{e}}^{\mu }=(e^{i},\mathbf{%
\check{e}}^{a})$ in explicit form by introducing $\check{N}_{\ \underline{i}%
}^{a^{\prime }},$ respectively, in formulas (\ref{dder}) and (\ref{ddif}).$%
\square $
\end{proof}

\vskip3pt Similar constructions can be defined for $\mathcal{L=F}^{2}(x,y),$
where an effective Finsler metric $\mathcal{F}$ is a differentiable function
of class $C^{\infty }$ in any point $(x,y)$ with $y\neq 0$ and is continuous
in any point $(x,0);$ $\mathcal{F}(x,y)>0$ if $y\neq 0;$ it satisfies the
homogeneity condition $\mathcal{F}(x,\beta y)=|\beta |\mathcal{F}(x,y)$ for
any nonzero $\beta \in \mathbb{R}$ and the Hessian (\ref{elf}) computed for $%
\mathcal{L=F}^{2}$ is positive definite. In this case, we can say that a
(pseudo) Riemannian space with metric $\mathbf{g}$ is modeled by an
effective Finsler geometry and, inversely, a Finsler geometry is modeled on
a (pseudo) Riemannian space. Such ideas were considered, for instance, in
Ref. \cite{ma} for Lagrange and Finsler spaces defined on tangent bundles.
In Ref. \cite{vrfg}, we model such geometries on (pseudo) Riemannian and
Riemann--Cartan spaces endowed with nonholonomic distributions.

\begin{definition}
A (pseudo) Riemannian manifold $\mathbf{V}^{2n}$ is nonholonomic
(N-\--an\-holonomic) if it is provided with a nonholonomic distribution on $%
TV^{2n}$ (N--connection structure $\mathbf{N}$).
\end{definition}

We formulate the first main result in this paper:

\begin{theorem}
\label{mth1}Any (pseudo) Riemannian space can be transformed into a
N--anho\-lo\-nomic manifold $\mathbf{V}^{2n}$ modeling an effective Lagrange
(or Finsler) geometry by prescribing a generating Lagrange (or Finsler)
function $\mathcal{L}(x,y)$ (or $\mathcal{F}(x,y)).$
\end{theorem}

\begin{proof}
Such a proof follows from Propositions \ref{pr01} and \ref{pr02}\ and Lemma %
\ref{lem01}. It should be noted that, by corresponding vielbein transforms $%
e_{\ \underline{i}}^{i},e_{\ \underline{i}}^{a}$ and $e_{\ \underline{a}%
}^{a},$ any metric $\mathbf{g}$ with coefficients defined with respect to an
arbitrary co--frame $\mathbf{e}^{\mu },$ see (\ref{m1}), can be transformed
into canonical Lagrange (Finsler) ones, $\mathbf{\check{g}}$ (\ref{hvmetr1}%
). The $\mathbf{\check{g}}$ coefficients are computed with respect to $%
\mathbf{\check{e}}^{\mu }=(e^{i},\mathbf{\check{e}}^{a}),$ with the
associated N--connection structure $\check{N}_{\ \underline{i}}^{a^{\prime
}},$ all defined by a prescribed $\mathcal{L}(x,y)$ (or $\mathcal{F}%
(x,y)).\square $
\end{proof}

\vskip3pt

Finally, it should be noted that considering an arbitrary effective
Lagrangian $\mathcal{L}(x,y)$ on a four dimensional (pseudo) Riemannian
spacetime and defining a corresponding $2+2$ decomposition, local Lorentz
invariance is not violated. We can work in any reference frame and
coordinates, but the constructions adapted to the canonical N--connection
structure and an analogous mechanical modeling are more convenient for
developing a formalism of deformation quantization of gravity following the
appropriate methods for Lagrange--Finsler and almost K\"{a}hler spaces.

\section{Almost K\"{a}hler Models for (Pseudo) Riemannian and Lagrange Spaces%
}

The goal of this section is to prove that for any (pseudo) Riemannian metric
and $n+n$ splitting we can define canonical almost symplectic structures.
The analogous mechanical modeling developed in previous sections is
important from two points of view: Firstly, it provides both geometric and
physical interpretations for the class of nonholonomic transforms with $n+n$
splitting and adapting to the N--connection. Secondly, such canonical
constructions can be equivalently redefined as a class of almost K\"{a}hler
geometries with associated N--connection when certain symplectic forms and
linear connection structures are canonically induced by the metric $\mathbf{g%
}(x,y)$ and effective Lagrangian $\mathcal{L}(x,y)$ on $\mathbf{V}^{2n}.$

\subsection{Canonical Riemann--Lagrange symplectic struc\-tu\-res}

Let $\mathbf{\check{e}}_{\alpha ^{\prime }}=(\mathbf{\check{e}}%
_{i},e_{b^{\prime }})$ and $\mathbf{\check{e}}^{\alpha ^{\prime }}=(e^{i},\
\mathbf{\check{e}}^{b^{\prime }})$ be defined respectively by (\ref{dder})
and (\ref{ddif}) for the canonical N--connection $\mathbf{\check{N}}$ stated
by a metric structure $\mathbf{g}=\mathbf{\check{g}}$ on $\mathbf{V}^{2n}.$
We introduce a linear operator $\mathbf{\check{J}}$ acting on tangent
vectors to $\mathbf{V}^{2n}$ following formulas $\mathbf{\check{J}}(\mathbf{%
\check{e}}_{i})=-e_{n+i}$\ and \ $\mathbf{\check{J}}(e_{n+i})=\mathbf{%
\check{e}}_{i},$ where the index $a^{\prime }$ runs values $n+i$ for $%
i=1,2,...n$ and $\mathbf{\check{J}\circ \check{J}=-I}$ for $\mathbf{I}$
being the unity matrix. Equivalently, we introduce a tensor field on $%
\mathbf{V}^{2n},$%
\begin{eqnarray*}
\mathbf{\check{J}} &=&\mathbf{\check{J}}_{\ \beta }^{\alpha }\ e_{\alpha
}\otimes e^{\beta }=\mathbf{\check{J}}_{\ \underline{\beta }}^{\underline{%
\alpha }}\ \frac{\partial }{\partial u^{\underline{\alpha }}}\otimes du^{%
\underline{\beta }} \\
&=&\mathbf{\check{J}}_{\ \beta ^{\prime }}^{\alpha ^{\prime }}\ \mathbf{%
\check{e}}_{\alpha ^{\prime }}\otimes \mathbf{\check{e}}^{\beta ^{\prime }}=%
\mathbf{-}e_{n+i}\otimes e^{i}+\mathbf{\check{e}}_{i}\otimes \ \mathbf{%
\check{e}}^{n+i} \\
&=&-\frac{\partial }{\partial y^{i}}\otimes dx^{i}+\left( \frac{\partial }{%
\partial x^{i}}-\check{N}_{i}^{n+j}\frac{\partial }{\partial y^{j}}\right)
\otimes \left( dy^{i}+\check{N}_{k}^{n+i}dx^{k}\right) .
\end{eqnarray*}%
It is clear that $\mathbf{\check{J}}$ defines globally an almost complex
structure on\ $\mathbf{V}^{2n}$ completely determined by a fixed $\mathcal{L}%
(x,y).$ Using vielbeins $\mathbf{e}_{\ \underline{\alpha }}^{\alpha }$ and
their duals $\mathbf{e}_{\alpha \ }^{\ \underline{\alpha }}$, defined by $%
e_{\ \underline{i}}^{i},e_{\ \underline{i}}^{a}$ and $e_{\ \underline{a}%
}^{a} $ as a solution of (\ref{aleq}), we can compute the coefficients $%
\mathbf{\check{J}} $ with respect to any local basis $e_{\alpha }$ and $%
e^{\alpha }$ on $\mathbf{V}^{n+n},$ $\mathbf{\check{J}}_{\ \beta }^{\alpha }=%
\mathbf{e}_{\ \underline{\alpha }}^{\alpha }\mathbf{\check{J}}_{\ \underline{%
\beta }}^{\underline{\alpha }}\mathbf{e}_{\beta \ }^{\ \underline{\beta }}.$
In general, we can define an almost complex structure $\mathbf{J}$ for an
arbitrary N--connection $\mathbf{N}$ by using N--adapted bases (\ref{dder})
and (\ref{ddif}), not necessarily induced by an effective Lagrange function.

\begin{definition}
The Nijenhuis tensor field for any almost complex structure $\mathbf{J}$
determined by a N--connection (equivalently, the curvature of
N--connecti\-on) is defined as
\begin{equation}
\ ^{\mathbf{J}}\mathbf{\Omega (X,Y)=-[X,Y]+[JX,JY]-J[JX,Y]-J[X,JY],}
\label{neijt}
\end{equation}%
for any d--vectors $\mathbf{X}$ and $\mathbf{Y.}$
\end{definition}

With respect to N--adapted bases (\ref{dder}) and (\ref{ddif}) the
Neijenhuis tensor $\ ^{\mathbf{J}}\mathbf{\Omega =\{}\Omega _{ij}^{a}\mathbf{%
\}}$ has the coefficients
\begin{equation}
\Omega _{ij}^{a}=\frac{\partial N_{i}^{a}}{\partial x^{j}}-\frac{\partial
N_{j}^{a}}{\partial x^{i}}+N_{i}^{b}\frac{\partial N_{j}^{a}}{\partial y^{b}}%
-N_{j}^{b}\frac{\partial N_{i}^{a}}{\partial y^{b}}.  \label{nccurv}
\end{equation}%
A N--anholonomic manifold $\mathbf{V}^{2n}$ is integrable if $\Omega
_{ij}^{a}=0.$ We get a complex structure if and only if both the h-- and
v--distributions are integrable, i.e., if and only if $\Omega _{ij}^{a}=0$
and $\frac{\partial N_{j}^{a}}{\partial y^{i}}-\frac{\partial N_{i}^{a}}{%
\partial y^{j}}=0.$

\begin{definition}
An almost symplectic structure on a manifold $V^{n+m},$ \newline
$\dim V^{n+m}=n+m,$ is defined by a nondegenerate 2--form
$\theta =\frac{1}{2}\theta _{\alpha \beta }(u)e^{\alpha }\wedge e^{\beta }.$
\end{definition}

We have

\begin{proposition}
For any $\theta $ on $V^{n+m},$ there is a unique N--connection $\mathbf{N}%
=\{N_{i}^{a}\}$ defined as a splitting $TV^{n+m}=hV^{n+m}\oplus vV^{n+m},$
where indices $i,j,..=1,2,...n$ and $a,b,...=n+1,n+1,...n+m$. The function $%
\theta $ satisfies the following conditions:%
\begin{equation}
\theta =(h\mathbf{X},v\mathbf{Y})=0\mbox{ and }\theta =h\theta +v\theta ,
\label{aux02}
\end{equation}%
for any $\mathbf{X}=h\mathbf{X}+v\mathbf{X,}$ $\mathbf{Y}=h\mathbf{Y}+v%
\mathbf{Y}$ and $h\theta (\mathbf{X,Y})\doteqdot \theta (h\mathbf{X,}h%
\mathbf{Y}),$\newline
$v\theta (\mathbf{X,Y})\doteqdot \theta (v\mathbf{X,}v\mathbf{Y}).$ Here the
symbol ''$\doteqdot $'' means ''by definition''.
\end{proposition}

\begin{proof}
For $\mathbf{X=e}_{\alpha }=(\mathbf{e}_{i},e_{a})$ and $\mathbf{Y=e}%
_{\beta}=(\mathbf{e}_{l},e_{b}),$ where $\mathbf{e}_{\alpha }$ is a
N--adapted basis\ of type (\ref{dder}) of dimension $n+m,$ we write the
first equation in (\ref{aux02}) as $\theta =\theta (\mathbf{e}%
_{i},e_{a})=\theta (\frac{\partial }{\partial x^{i}},\frac{\partial }{%
\partial y^{a}})-N_{i}^{b}\theta (\frac{\partial }{\partial y^{b}},\frac{%
\partial }{\partial y^{a}})=0.$  We can find a unique solution form and
define $N_{i}^{b}$ if $rank|\theta (\frac{\partial }{\partial y^{b}},\frac{%
\partial }{\partial y^{a}})|=m.$ Denoting locally
\begin{equation}
\theta =\frac{1}{2}\theta _{ij}(u)e^{i}\wedge e^{j}+\frac{1}{2}\theta
_{ab}(u)\mathbf{e}^{a}\wedge \mathbf{e}^{b},  \label{aux03}
\end{equation}%
where the first term is for $h\theta $ and the second term is $v\theta ,$ we
get the second formula in (\ref{aux02}). We may consider the particular case
in which $n=m.\square $
\end{proof}

\begin{definition}
An almost Hermitian model of a (pseudo) Riemannian spa\-ce $\mathbf{V}^{2n}$
equipped with an N--connection structure $\mathbf{N}$ is defined by a triple
$\mathbf{H}^{2n}=(\mathbf{V}^{2n},\theta ,\mathbf{J}),$ where $\mathbf{%
\theta (X,Y)}\doteqdot \mathbf{g}\left( \mathbf{JX,Y}\right).$
\end{definition}

In addition, we have

\begin{definition}
A space $\mathbf{H}^{2n}$ is almost K\"{a}hler, denoted $\mathbf{K}^{2n},$
if and only if $d\mathbf{\theta }=0.$
\end{definition}

If a (pseudo) Riemannian space is modeled by a Lagrange--Finsler geometry,
the second main result of this paper follows

\begin{theorem}
\label{thmr2}Having chosen a generating function $\mathcal{L}(x,y)$ (or $%
\mathcal{F}(x,y))$ on a (pseudo) Riemannian manifold $V^{n+n},$ we can model
this space as an almost K\"{a}hler geometry, i.e. $\mathbf{\check{H}}^{2n}=%
\mathbf{\check{K}}^{2n}.$
\end{theorem}

\begin{proof}
For $\mathbf{g}=\mathbf{\check{g}}$ (\ref{hvmetr1}) and structures $\mathbf{%
\check{N}}$ and $\mathbf{\check{J}}$ canonically defined by $\mathcal{L},$
we define $\mathbf{\check{\theta}(X,Y)}\doteqdot \mathbf{\check{J}}\left(
\mathbf{\check{F}X,Y}\right) $ for any d--vectors $\mathbf{X}$ and $\mathbf{%
Y.}$ In local N--adapted form form, we have
\begin{eqnarray}
\mathbf{\check{\theta}} &=&\frac{1}{2}\check{\theta}_{\alpha \beta
}(u)e^{\alpha }\wedge e^{\beta }=\frac{1}{2}\check{\theta}_{\underline{%
\alpha }\underline{\beta }}(u)du^{\underline{\alpha }}\wedge du^{\underline{%
\beta }}  \label{asymstr} \\
&=&\check{g}_{ij}(x,y)\check{e}^{n+i}\wedge dx^{j}=\check{g}%
_{ij}(x,y)(dy^{n+i}+\check{N}_{k}^{n+i}dx^{k})\wedge dx^{j}.  \notag
\end{eqnarray}%
Let us consider the form $\check{\omega}=\frac{1}{2}\frac{\partial \mathcal{L%
}}{\partial y^{n+i}}dx^{i}.$ A straightforward computation, using
Proposition \ref{pr02} and N--connection $\mathbf{\check{N}}$ (\ref{ncel}),
shows that $\mathbf{\check{\theta}}=d\check{\omega},$ which means that $d%
\mathbf{\check{\theta}}=dd\check{\omega}=0$ and that the canonical effective
Lagrange structures $\mathbf{g}=\mathbf{\check{g},\check{N}}$ and $\mathbf{%
\check{J}}$ induce an almost K\"{a}hler geometry. Instead of "Lagrangian
mechanics variables" we can introduce another type redefining $\mathbf{%
\check{\theta}}$ with respect to an arbitrary co--frame basis using
vielbeins $\mathbf{e}_{\ \underline{\alpha }}^{\alpha } $ and their duals $%
\mathbf{e}_{\alpha \ }^{\ \underline{\alpha }},$ defined by $e_{\ \underline{%
i}}^{i},e_{\ \underline{i}}^{a}$ and $e_{\ \underline{a}}^{a}$ (\ref{aleq}).
So, we can compute $\check{\theta}_{\alpha \beta }=\mathbf{e}_{\alpha \ }^{\
\underline{\alpha }}\mathbf{e}_{\beta \ }^{\ \underline{\beta }}\check{\theta%
}_{\underline{\alpha }\underline{\beta }}$ and express the 2--form (\ref%
{asymstr})as $\check{\theta}=\frac{1}{2}\check{\theta}_{ij}(u)e^{i}\wedge
e^{j}+\frac{1}{2}\check{\theta}_{ab}(u)\mathbf{\check{e}}^{a}\wedge \mathbf{%
\check{e}}^{b},$ see (\ref{aux03}). The coefficients $\check{\theta}_{ab}=%
\check{\theta}_{n+i\ n+j}$ above are equal, respectively, to the
coefficients $\check{\theta}_{ij}$ and the dual N--adapted basis $\mathbf{%
\check{e}}^{\alpha }=(e^{i},\mathbf{\check{e}}^{a})$ is elongated by $%
\check{N}_{j}^{a}$ (\ref{ncel}). It should be noted that for a general
2--form $\theta $ directly constructed from  a metric $\mathbf{g}$ and
almost complex $\mathbf{J}$\textbf{\ }structures on $V^{2n}$, we have that $%
d\theta \neq 0.$ For a $n+n$ splitting induced by an effective Lagrange
(Finsler) generating function, we have $d\mathbf{\check{\theta}}=0$ which
results in a canonical almost K\"{a}hler model completely defined by $%
\mathbf{g}=\mathbf{\check{g}}$ and chosen $\mathcal{L}(x,y)$ (or $\mathcal{F}%
(x,y)).$ $\square $
\end{proof}

\subsection{N--adapted symplectic connections}

In our approach, we work with nonholonomic (pseudo) Riemannian manifolds $%
\mathbf{V}^{2n}$ enabled with an effective N--connection and almost
symplectic structures defined canonically by the metric structure $\mathbf{g}%
=\mathbf{\check{g}}$ and a fixed $\mathcal{L}(x,y).$ In this section, we
analyze the class of linear connections that can be adapted to the
N--connection and/or symplectic structure and defined canonically if a
corresponding nonholonomic distribution is induced completely by $\mathcal{L}%
,$ or $\mathcal{F}.$

From the class of arbitrary affine connections on $\mathbf{V}^{2n},$ one
prefers to work with N--adapted linear connections, called distinguished
connections ( d--connections).

\begin{definition}
A linear connection on $\mathbf{V}^{2n}$ is a d--connection
\begin{equation*}
\mathbf{D}=(hD;vD)=\{\mathbf{\Gamma }_{\beta \gamma }^{\alpha
}=(L_{jk}^{i},\ ^{v}L_{bk}^{a};C_{jc}^{i},\ ^{v}C_{bc}^{a})\},
\end{equation*}%
with local coefficients computed with respect to (\ref{dder}) and (\ref{ddif}%
), which preserves the distribution (\ref{whitney}) under parallel
transports.
\end{definition}

For a d--connection $\mathbf{D,}$ we can define respectively the torsion and
curvature tensors,
\begin{eqnarray}
\mathbf{T}(\mathbf{X},\mathbf{Y}) &\doteqdot &\mathbf{D}_{\mathbf{X}}\mathbf{%
Y}-\mathbf{D}_{\mathbf{Y}}\mathbf{X}-[\mathbf{X},\mathbf{Y}],  \label{ators}
\\
\mathbf{R}(\mathbf{X},\mathbf{Y})\mathbf{Z} &\doteqdot &\mathbf{D}_{\mathbf{X%
}}\mathbf{D}_{\mathbf{Y}}\mathbf{Z}-\mathbf{D}_{\mathbf{Y}}\mathbf{D}_{%
\mathbf{X}}\mathbf{Z}-\mathbf{D}_{[\mathbf{X},\mathbf{Y}]}\mathbf{Z},
\label{acurv}
\end{eqnarray}%
where $[\mathbf{X},\mathbf{Y}]\doteqdot \mathbf{XY}-\mathbf{YX,}$ for any
d--vectors $\mathbf{X} $ and $\mathbf{Y}.$ The coefficients $\mathbf{T}=\{%
\mathbf{T}_{\ \beta \gamma }^{\alpha }\}$ and $\mathbf{R}=\{\mathbf{R}_{\
\beta \gamma \tau }^{\alpha }\}$ can be written in terms of $\mathbf{e}%
_{\alpha }$ and $\mathbf{e}^{\beta }$ by introducing $\mathbf{X}\rightarrow
\mathbf{e}_{\alpha },\mathbf{Y}\rightarrow \mathbf{e}_{\beta }\mathbf{,Z}%
\rightarrow \mathbf{e}_{\gamma }$ in (\ref{ators}) and (\ref{acurv}), see
Ref. \cite{vrfg} for details.

\begin{definition}
A d--connection $\mathbf{D}$\ is metric compatible with a d--metric $\mathbf{%
g}$ if $\mathbf{D}_{\mathbf{X}}\mathbf{g}=0$ for any d--vector field $%
\mathbf{X.}$
\end{definition}

If an almost symplectic structure is defined on a N--anholonomic manifold,
one considers:

\begin{definition}
\label{defasstr}An almost symplectic d--connection $\ _{\theta }\mathbf{D}$
on $\mathbf{V}^{2n},$ or (equivalently) a d--connection compatible with an
almost symplectic structure $\theta ,$ is defined such that $\ _{\theta }%
\mathbf{D}$ is N--adapted, i.e., it is a d--connection, and $\ _{\theta }%
\mathbf{D}_{\mathbf{X}}\theta =0,$ for any d--vector $\mathbf{X.}$
\end{definition}

We can always fix a d--connection $\ _{\circ }\mathbf{D}$ on $\mathbf{V}%
^{2n} $ and then construct an almost symplectic $\ _{\theta }\mathbf{D.}$

\begin{example}
Let us represent $\theta $ in N--adapted form (\ref{aux03}). Having chosen a
\begin{eqnarray*}
\ _{\circ }\mathbf{D} &=&\left\{ h\ _{\circ }D=(\ _{\circ }D_{k},\ \ _{\circ
}^{v}D_{k});v\ _{\circ }D=(\ _{\circ }D_{c},\ \ _{\circ }^{v}D_{c})\right\}
\\
&=&\{\ _{\circ }\mathbf{\Gamma }_{\beta \gamma }^{\alpha }=(\ _{\circ
}L_{jk}^{i},\ _{\circ }^{v}L_{bk}^{a};\ _{\circ }C_{jc}^{i},\ _{\circ
}^{v}C_{bc}^{a})\},
\end{eqnarray*}%
we can verify that
\begin{eqnarray*}
\ _{\theta }\mathbf{D} &=&\left\{ h\ _{\theta }D=(\ _{\theta }D_{k},\ \
_{\theta }^{v}D_{k});v\ _{\theta }D=(\ _{\theta }D_{c},\ \ _{\theta
}^{v}D_{c})\right\} \\
&=&\{\ _{\theta }\mathbf{\Gamma }_{\beta \gamma }^{\alpha }=(\ _{\theta
}L_{jk}^{i},\ _{\theta }^{v}L_{bk}^{a};\ _{\theta }C_{jc}^{i},\ _{\theta
}^{v}C_{bc}^{a})\},
\end{eqnarray*}%
with
\begin{eqnarray}
\ _{\theta }L_{jk}^{i} &=&\ _{\circ }L_{jk}^{i}+\frac{1}{2}\theta ^{ih}\
_{\circ }D_{k}\theta _{jh},\ \ _{\theta }^{v}L_{bk}^{a}=\ _{\circ
}^{v}L_{bk}^{a}+\frac{1}{2}\theta ^{ae}\ _{\circ }^{v}D_{k}\theta _{eb},
\label{csdc} \\
\ _{\theta }C_{jc}^{i} &=&\ _{\theta }C_{jc}^{i}+\frac{1}{2}\theta ^{ih}\
_{\circ }D_{c}\theta _{jh},\ \ _{\theta }^{v}C_{bc}^{a}=\ _{\circ
}^{v}C_{bc}^{a}+\frac{1}{2}\theta ^{ae}\ _{\circ }^{v}D_{c}\theta _{eb},
\notag
\end{eqnarray}%
satisfies the conditions $\ _{\theta }D_{k}\theta _{jh}=0,\ \ _{\theta
}^{v}D_{k}\theta _{eb}=0,\ _{\theta }D_{c}\theta _{jh}=0,\ _{\theta
}^{v}D_{c}\theta _{eb}=0,$  which is equivalent to $\ _{\theta }\mathbf{D}_{%
\mathbf{X}}\theta =0$ from Definition \ref{defasstr}.
\end{example}

Let us introduce the operators
\begin{equation}
\Theta _{jk}^{hi}=\frac{1}{2}(\delta _{j}^{h}\delta _{k}^{i}-\theta
_{jk}\theta ^{ih})\mbox{ and }\Theta _{cd}^{ab}=\frac{1}{2}(\delta
_{c}^{a}\delta _{d}^{b}-\theta _{cd}\theta ^{ab}),  \label{thop}
\end{equation}%
with the coefficients computed with respect to N--adapted bases (\ref{dder})
and (\ref{ddif}). By straightforward computations, one proves the following
theorem.

\begin{theorem}
The set of d--connections $\ _{s}\mathbf{\Gamma }_{\beta \gamma }^{\alpha
}=(\ _{s}L_{jk}^{i},\ _{s}^{v}L_{bk}^{a};\ _{s}C_{jc}^{i},\
_{s}^{v}C_{bc}^{a})$, compatible with an almost symplectic structure $\theta
$ (\ref{aux03}), are parametrized by
\begin{eqnarray}
\ _{s}L_{jk}^{i} &=&\ _{\theta }L_{jk}^{i}+\Theta _{jl}^{hi}\ Y_{hk}^{l},\
_{s}^{v}L_{bk}^{a}=\ _{\theta }^{v}L_{bk}^{a}+\Theta _{bd}^{ca}\ Y_{ck}^{d},
\label{fsdc} \\
\ _{s}C_{jc}^{i} &=&\ _{\theta }C_{jc}^{i}+\Theta _{jl}^{hi}\ Y_{hc}^{l},\
_{s}^{v}C_{bc}^{a}=\ _{\theta }^{v}C_{bc}^{a}+\Theta _{bd}^{ea}\ Y_{ec}^{d},
\notag
\end{eqnarray}%
where $\ _{\theta }\mathbf{\Gamma }_{\beta \gamma }^{\alpha }=(\ _{\theta
}L_{jk}^{i},\ _{\theta }^{v}L_{bk}^{a};\ _{\theta }C_{jc}^{i},\ _{\theta
}^{v}C_{bc}^{a})$ is given by (\ref{csdc}), the $\Theta $--operators are
those from (\ref{thop}) and $\mathbf{Y}_{\beta \gamma }^{\alpha }=\left(
Y_{jk}^{i},Y_{bk}^{a},Y_{jc}^{i},Y_{bc}^{a}\right) $ are arbitrary d--tensor
fields.
\end{theorem}

From the set of metric and/or almost symplectic compatible d--connecti\-ons
on a (pseudo) Riemannian manifold $V^{2n},$ we can select those which are
completely defined by $\mathbf{g}$ and a prescribed effective Lagrange
structure $\mathcal{L}(x,y):$

\begin{theorem}
There is a unique normal d--connection
\begin{eqnarray}
\ \widehat{\mathbf{D}} &=&\left\{ h\widehat{D}=(\widehat{D}_{k},^{v}\widehat{%
D}_{k}=\widehat{D}_{k});v\widehat{D}=(\widehat{D}_{c},\ ^{v}\widehat{D}_{c}=%
\widehat{D}_{c})\right\}  \label{ndc} \\
&=&\{\widehat{\mathbf{\Gamma }}_{\beta \gamma }^{\alpha }=(\widehat{L}%
_{jk}^{i},\ ^{v}\widehat{L}_{n+j\ n+k}^{n+i}=\widehat{L}_{jk}^{i};\ \widehat{%
C}_{jc}^{i}=\ ^{v}\widehat{C}_{n+j\ c}^{n+i},\ ^{v}\widehat{C}_{bc}^{a}=%
\widehat{C}_{bc}^{a})\},  \notag
\end{eqnarray}%
which is metric compatible, $\widehat{D}_{k}\check{g}_{ij}=0$ and $\widehat{D%
}_{c}\check{g}_{ij}=0,$ and completely defined by $\mathbf{g}=\mathbf{\check{%
g}}$ and a fixed $\mathcal{L}(x,y).$
\end{theorem}

\begin{proof}
First, we note that if a normal d--connection exists, it is completely
defined by couples of h-- and v--components $\ \widehat{\mathbf{D}}_{\alpha
}=(\widehat{D}_{k},\widehat{D}_{c}),$ i.e. $\widehat{\mathbf{\Gamma }}%
_{\beta \gamma }^{\alpha }=(\widehat{L}_{jk}^{i},\ ^{v}\widehat{C}%
_{bc}^{a}). $ Choosing
\begin{equation}
\widehat{L}_{jk}^{i}=\frac{1}{2}\check{g}^{ih}\left( \mathbf{\check{e}}_{k}%
\check{g}_{jh}+\mathbf{\check{e}}_{j}\check{g}_{hk}-\mathbf{\check{e}}_{h}%
\check{g}_{jk}\right) ,\widehat{C}_{jk}^{i}=\frac{1}{2}\check{g}^{ih}\left(
\frac{\partial \check{g}_{jh}}{\partial y^{k}}+\frac{\partial \check{g}_{hk}%
}{\partial y^{j}}-\frac{\partial \check{g}_{jk}}{\partial y^{h}}\right) ,
\label{cdcc}
\end{equation}%
where $\mathbf{\check{e}}_{k}=\partial /\partial x^{k}+\check{N}%
_{k}^{a}\partial /\partial y^{a},$ $\check{N}_{k}^{a}$ and $\check{g}_{jk}=%
\check{h}_{n+i\ n+j}$ are defined by canonical values (\ref{elf}) and (\ref%
{ncel}) induced by a regular $\mathcal{L}(x,y),$ we can prove that this
d--connection is unique and satisfies the conditions of the theorem. \ Using
vielbeins $\mathbf{e}_{\ \underline{\alpha }}^{\alpha }$ and their duals $%
\mathbf{e}_{\alpha \ }^{\ \underline{\alpha }},$ defined by $e_{\ \underline{%
i}}^{i},e_{\ \underline{i}}^{a}$ and $e_{\ \underline{a}}^{a}$ satisfying (%
\ref{aleq}), we can compute the coefficients of $\widehat{\mathbf{\Gamma }}%
_{\beta \gamma }^{\alpha }$ (\ref{ndc}) with respect to arbitrary frame
basis $e_{\alpha }$ and co--basis $e^{\alpha }$ on $V^{n+m}.\square $
\end{proof}

\vskip5pt Introducing the normal d--connection 1--form $\widehat{\mathbf{%
\Gamma }}_{j}^{i}=\widehat{L}_{jk}^{i}e^{k}+\widehat{C}_{jk}^{i}\mathbf{%
\check{e}}^{k},$ for $e^{k}=dx^{k}$ and $\mathbf{\check{e}}^{k}=dy^{k}+%
\check{N}_{i}^{k}dx^{k},$ we can prove that the Cartan structure equations
are satisfied,%
\begin{equation}
de^{k}-e^{j}\wedge \widehat{\mathbf{\Gamma }}_{j}^{k}=-\widehat{\mathcal{T}}%
^{i},\ d\mathbf{\check{e}}^{k}-\mathbf{\check{e}}^{j}\wedge \widehat{\mathbf{%
\Gamma }}_{j}^{k}=-\ ^{v}\widehat{\mathcal{T}}^{i},  \label{cart1}
\end{equation}%
and
\begin{equation}
d\widehat{\mathbf{\Gamma }}_{j}^{i}-\widehat{\mathbf{\Gamma }}_{j}^{h}\wedge
\widehat{\mathbf{\Gamma }}_{h}^{i}=-\widehat{\mathcal{R}}_{\ j}^{i}.
\label{cart2}
\end{equation}

The h-- and v--components of the torsion 2--form $\widehat{\mathcal{T}}%
^{\alpha }=\left( \widehat{\mathcal{T}}^{i},\ ^{v}\widehat{\mathcal{T}}%
^{i}\right) =\widehat{\mathbf{T}}_{\ \tau \beta}^{\alpha }\ \mathbf{\check{e}%
}^{\tau }\wedge \mathbf{\check{e}}^{\beta }$  and from (\ref{cart1}) the
components are computed
\begin{equation}
\widehat{\mathcal{T}}^{i}=\widehat{C}_{jk}^{i}e^{j}\wedge \mathbf{\check{e}}%
^{k},\ ^{v}\widehat{\mathcal{T}}^{i}=\frac{1}{2}\check{\Omega}%
_{kj}^{i}e^{k}\wedge e^{j}+(\frac{\partial \check{N}_{k}^{i}}{\partial y^{j}}%
-\widehat{L}_{\ kj}^{i})e^{k}\wedge \mathbf{\check{e}}^{j},  \label{tform}
\end{equation}%
where $\check{\Omega}_{kj}^{i}$ are coefficients of the curvature of the
canonical N--connection $\check{N}_{k}^{i}$ defined by formulas similar to (%
\ref{nccurv}). Such formulas also follow from (\ref{ators}) redefined for $%
\widehat{\mathbf{D}}_{\alpha }$ and $\mathbf{\check{e}}_{\alpha },$ when the
torsion $\widehat{\mathbf{T}}_{\beta \gamma }^{\alpha }$ is parametrized as
\begin{equation}
\widehat{T}_{jk}^{i}=0,\widehat{T}_{jc}^{i}=\widehat{C}_{\ jc}^{i},\widehat{T%
}_{ij}^{a}=\check{\Omega}_{ij}^{a},\widehat{T}_{ib}^{a}=e_{b}\check{N}%
_{i}^{a}-\widehat{L}_{\ bi}^{a},\widehat{T}_{bc}^{a}=0.  \label{cdtors}
\end{equation}%
It should be noted that $\widehat{\mathbf{T}}$ vanishes on h- and
v--subspaces, i.e. $\widehat{T}_{jk}^{i}=0$ and $\widehat{T}_{bc}^{a}=0,$
but certain nontrivial h--v--components induced by the nonholonomic
structure are defined canonically by $\mathbf{g}=\mathbf{\check{g}}$ and $%
\mathcal{L}.$

We can also compute the curvature 2--form from (\ref{cart2}),%
\begin{equation}
\widehat{\mathcal{R}}_{\ \gamma }^{\tau }=\widehat{\mathbf{R}}_{\ \gamma
\alpha \beta }^{\tau }\ \mathbf{\check{e}}^{\alpha }\wedge \ \mathbf{%
\check{e}}^{\beta }=\frac{1}{2}\widehat{R}_{\ jkh}^{i}e^{k}\wedge e^{h}+%
\widehat{P}_{\ jka}^{i}e^{k}\wedge \mathbf{\check{e}}^{a}+\frac{1}{2}\
\widehat{S}_{\ jcd}^{i}\mathbf{\check{e}}^{c}\wedge \mathbf{\check{e}}^{d},
\label{cform}
\end{equation}%
where the nontrivial N--adapted coefficients of curvature $\widehat{\mathbf{R%
}}_{\ \beta \gamma \tau }^{\alpha }$ of $\widehat{\mathbf{D}}$ are (such
formulas can be proven also from (\ref{acurv}) written for $\widehat{\mathbf{%
D}}_{\alpha }$ and $\mathbf{\check{e}}_{\alpha })$
\begin{eqnarray}
\widehat{R}_{\ hjk}^{i} &=&\mathbf{\check{e}}_{k}\widehat{L}_{\ hj}^{i}-%
\mathbf{\check{e}}_{j}\widehat{L}_{\ hk}^{i}+\widehat{L}_{\ hj}^{m}\widehat{L%
}_{\ mk}^{i}-\widehat{L}_{\ hk}^{m}\widehat{L}_{\ mj}^{i}-\widehat{C}_{\
ha}^{i}\check{\Omega}_{\ kj}^{a},  \label{cdcurv} \\
\widehat{P}_{\ jka}^{i} &=&e_{a}\widehat{L}_{\ jk}^{i}-\widehat{\mathbf{D}}%
_{k}\widehat{C}_{\ ja}^{i},\ \widehat{S}_{\ bcd}^{a}=e_{d}\widehat{C}_{\
bc}^{a}-e_{c}\widehat{C}_{\ bd}^{a}+\widehat{C}_{\ bc}^{e}\widehat{C}_{\
ed}^{a}-\widehat{C}_{\ bd}^{e}\widehat{C}_{\ ec}^{a}.  \notag
\end{eqnarray}%
If instead of an effective Lagrange function one considers a Finsler
generating fundamental function $\mathcal{F}^{2},$ similar formulas for the
torsion and curvature of the normal d--connection can also be found.

There is another very important property of the normal d--connection:

\begin{theorem}
The normal d--connection $\widehat{\mathbf{D}}$ defines a unique almost
symplectic d--connection, $\widehat{\mathbf{D}}\equiv \ _{\theta }\widehat{%
\mathbf{D}},$ see Definition \ref{defasstr}, which is N--adapted, i.e. it
preserves under parallelism the splitting (\ref{whitney}), $_{\theta }%
\widehat{\mathbf{D}}_{\mathbf{X}}\check{\theta}\mathbf{=}0$ and $\widehat{T}%
_{jk}^{i}=\widehat{T}_{bc}^{a}=0,$ i.e. the torsion is of type (\ref{cdtors}%
).
\end{theorem}

\begin{proof}
Applying the conditions of the theorem to the coefficients (\ref{cdcc}), the
proof follows in a straightforward manner. $\square $
\end{proof}

\vskip3pt

It is pertinent to note that the normal d--connection $\widehat{\mathbf{%
\Gamma }}_{\beta \gamma }^{\alpha }$ (\ref{ndc}) is a N--anholonomic analog
of the affine connection $\ ^{K}\mathbf{\Gamma }_{\beta \gamma }^{\alpha }$
and Nijenhuis tensor $^{K}\mathbf{\Omega }_{\ \beta \gamma }^{\alpha }$ with
the torsion satisfying the condition $\ ^{K}\mathbf{T}_{\ \beta \gamma
}^{\alpha }=(1/4)^{K}\mathbf{\Omega }_{\ \beta \gamma }^{\alpha },$ as
considered in Ref. \cite{karabeg1}. For trivial N--connection structures, by
corresponding frame and coordinate transforms, we can identify $\widehat{%
\mathbf{\Gamma }}_{\beta \gamma }^{\alpha }$ with $\ ^{K}\mathbf{\Gamma }%
_{\beta \gamma }^{\alpha }$ (we used this property in our former works \cite%
{vqgr1,vqgr2,vqgr3,vbrane}).

In this section, we proved that a N--adapted and almost symplectic $\widehat{%
\mathbf{\Gamma }}_{\beta \gamma }^{\alpha }$ can be uniquely defined by a
(pseudo) Riemannian metric $\mathbf{g}$ if we prescribe an effective
Lagrange, or Finsler, function $\mathcal{L},$ or $\mathcal{F}$ on $V^{2n}.$
This allows us to construct an analogous Lagrange model for gravity and, at
the next step, to transform it equivalently in an almost K\"{a}hler
structure adapted to a corresponding $n+n$ spacetime splitting. For the
Einstein metrics, we get a canonical $2+2$ decomposition for which we can
apply the Fedosov's quantization if the geometric objects and operators are
adapted to the associated N--connection.

\begin{definition}
A (pseudo) Riemannian space is described in Lagrange--Finsler variables if
its vielbein, metric and linear connection structures are equivalently
transformed into corresponding canonical N---connection,
La\-gran\-ge--Finsler metric and normal / almost symplectic d--connection
structures.
\end{definition}

It should be noted that former approaches to the canonical and quantum loop
quantization of gravity were elaborated for $3+1$ fibrations and
corresponding ADM and Ashtekar variables with further modifications. On the
other hand, in order to elaborate certain approaches to deformation
quantization of gravity, it is crucial to work with nonholonomic $2+2$
structures, which is more convenient for certain Lagrange geometrized
constructions and their almost symplectic variants. For other models, the $%
3+1$ splitting preserves a number of similarities to Hamilton mechanics. In
our approach, the spacetime decompositions are defined by corresponding
N--connection structures, which can be induced canonically by effective
Lagrange, or Finsler, generating functions. One works both with N--adapted
metric coefficients and nonholonomic frame coefficients, the last ones being
defined by generic off--diagonal metric coefficients and related
N--connection coefficients. In the models related to $3+1$ fibrations, one
works with shift functions and frame variables which contain all dynamical
information, instead of metrics.

We also discuss here the similarities and differences of preferred classes
of linear connections used for $3+1$ and $2+2$ structures. In the first
case, the Ashtekar variables (and further modifications) were proved to
simplify the constraint structure of a gauge like theory to which the
Einstein theory was transformed in order to develop a background independent
quantization of gravity. In the second case, the analogs of Ashtekar
variables are generated by a canonical Lagrange--Finsler type metric and/or
corresponding almost symplectic structure, both adapted to the N--connection
structure. It is also involved the normal d--connection which is compatible
with the almost symplectic structure and completely defined by the metric
structure, alternatively to the Levi--Civita connection (the last one is not
adapted to the N--connection and induced almost symplectic structure). In
fact, all constructions for the normal d--connection can be redefined in an
equivalent form to the Levi--Civita connection (see below section \ref%
{ssensp} and Refs. \cite{vsgg,vrfg}), or in Ashtekar variables, but in such
cases the canonical $2+2$ splitting and almost K\"{a}hler structure are
mixed by general frame and linear connection deformations.

Finally, it should be noted that in our approach we are inspired by a number
of results and methods from Finsler and Lagrange geometry. For instance, the
original proofs that Finsler and Lagrange geometries are equivalent to
certain classes of almost K\"{a}hler geometries with N--connection
structures were obtained in Refs. \cite{mats,opr1}, see also reviews \cite%
{ma1987,ma,vrfg}. But those constructions were elaborated for tangent
bundles which are not related to standard models of modern physics.
Re--defining the constructions for nonholonomic structures on classical and
quantum spacetime models, we could develop new important and effective
methods from the geometry of nonholonomic manifolds and apply them to
deformation quantization of gravity.

\section{Distinguished Fedosov's Operators}

The Fedosov's approach to deformation quantization \cite{fedos1,fedos2} will
be extended for (pseudo) Riemannian manifolds $V^{2n}$ endowed with an
effective Lagrange function $\mathcal{L}.$ The constructions elaborated in
Ref. \cite{karabeg1} will be adapted to the canonical N--connection
structure by considering decompositions with respect to $\mathbf{\breve{e}}%
_{\nu }=(\mathbf{\breve{e}}_{i},e_{a^{\prime }})$ and $\mathbf{\breve{e}}%
^{\mu }=(e^{i},\mathbf{\breve{e}}^{a^{\prime }})$ defined by a metric $%
\mathbf{g}$ (\ref{m1}). For simplicity, we shall work only with the normal/
almost symplectic d--connection, $\widehat{\mathbf{D}}\equiv \ _{\theta }%
\widehat{\mathbf{D}}$ (\ref{ndc}), see Definition \ref{defasstr}, but it
should be emphasized here that we can use any d--connection from the family (%
\ref{fsdc}) and develop a corresponding deformation quantization. Usually,
the proofs referring to constructions not adapted to N--connections \cite%
{karabeg1}, and on Lagrange (Finsler) spaces related to quantum gravity
models on tangent bundles \cite{vqgr2} will be sketched, while the details
can be found in the corresponding references. In this work, the formulas are
redefined on nonholonomic (pseudo) Riemannian manifolds modeling effective
regular mechanical systems and corresponding almost K\"{a}hler structures.

We introduce the tensor $\ \mathbf{\check{\Lambda}}^{\alpha \beta }\doteqdot
\check{\theta}^{\alpha \beta }-i\ \mathbf{\check{g}}^{\alpha \beta },$ where
$\check{\theta}^{\alpha \beta }$ is the form (\ref{asymstr}) with ''up''
indices and $\ \mathbf{\check{g}}^{\alpha \beta }$ is the inverse to $%
\mathbf{\check{g}}_{\alpha \beta }$ stated by coefficients of (\ref{hvmetr1}%
). The local coordinates on $\mathbf{V}^{2n}$ are parametrized as $%
u=\{u^{\alpha }\}$ and the local coordinates on $T_{u}\mathbf{V}^{2n}$ are
labeled $(u,z)=(u^{\alpha },z^{\beta }),$ where $z^{\beta }$ are fiber
coordinates.

The formalism of deformation quantization can be developed by using $%
C^{\infty }(V)[[v]]$, the space of formal series of variable $v$ with
coefficients from $C^{\infty }(V)$ on a Poisson manifold $(V,\{\cdot ,\cdot
\})$ (in this work, we deal with an almost Poisson structure defined by the
canonical almost symplectic structure). One defines an associative algebra
structure on $C^{\infty }(V)[[v]]$ with a $v$--linear and $v$--adically
continuous star product
\begin{equation}
\ ^{1}f\ast \ ^{2}f=\sum\limits_{r=0}^{\infty }\ _{r}C(\ ^{1}f,\ ^{2}f)\
v^{r},  \label{starp}
\end{equation}%
where $\ _{r}C,r\geq 0,$ are bilinear operators on $C^{\infty }(V)$ with $\
_{0}C(\ ^{1}f,\ ^{2}f)=\ ^{1}f\ ^{2}f$ and $\ _{1}C(\ ^{1}f,\ ^{2}f)-\
_{1}C(\ ^{2}f,\ ^{1}f)=i\{\ ^{1}f,\ ^{2}f\};$\ $i$ being the complex unity.
Constructions of type (\ref{starp}) are used for stating a formal Wick
product
\begin{equation}
a\circ b\ (z)\doteqdot \exp \left( i\frac{v}{2}\ \mathbf{\check{\Lambda}}%
^{\alpha \beta }\frac{\partial ^{2}}{\partial z^{\alpha }\partial
z_{[1]}^{\beta }}\right) a(z)b(z_{[1]})\mid _{z=z_{[1]}},  \label{fpr}
\end{equation}%
for two elements $a$ and $b$ defined by series of type
\begin{equation}
a(v,z)=\sum\limits_{r\geq 0,|\{\alpha \}|\geq 0}\ a_{r,\{\alpha
\}}(u)z^{\{\alpha \}}\ v^{r},  \label{formser}
\end{equation}%
where by $\{\alpha \}$ we label a multi--index. This way, we define a formal
Wick algebra $\mathbf{\check{W}}_{u}$ associated with the tangent space $%
T_{u}\mathbf{V}^{2n},$ for $u\in \mathbf{V}^{2n}.$ It should be noted that
the fibre product (\ref{fpr}) can be trivially extended to the space of $%
\mathbf{\check{W}}$--valued N--adapted differential forms $\mathcal{\check{W}%
}\otimes \Lambda $ by means of the usual exterior product of the scalar
forms $\Lambda ,$ where $\ \mathcal{\check{W}}$ denotes the sheaf of smooth
sections of $\mathbf{\check{W}.}$ There is a standard grading on $\Lambda $
denoted $\deg _{a}.$ One also introduces gradings $\deg _{v},\deg _{s},\deg
_{a}$ on $\ \mathcal{W}\otimes \Lambda $ defined on homogeneous elements $%
v,z^{\alpha },\mathbf{\check{e}}^{\alpha }$ as follows: $\deg _{v}(v)=1,$ $%
\deg _{s}(z^{\alpha })=1,$ $\deg _{a}(\mathbf{\check{e}}^{\alpha })=1,$ and
all other gradings of the elements $v,z^{\alpha },\mathbf{\check{e}}^{\alpha
}$ are set to zero. In this case, the product $\circ $ from (\ref{fpr}) on $%
\ \mathcal{\check{W}}\otimes \mathbf{\Lambda }$ is bigraded. This is written
w.r.t the grading $Deg=2\deg _{v}+\deg _{s}$ and the grading $\deg _{a}.$

\subsection{Normal Fedosov's d--operators}

The normal d--connection $\widehat{\mathbf{D}}\mathbf{=\{}\widehat{\mathbf{%
\Gamma }}\mathbf{_{\alpha \beta }^{\gamma }\}}$ (\ref{ndc}) can be extended
to operators
\begin{equation}
\widehat{\mathbf{D}}\left( a\otimes \lambda \right) \doteqdot \left( \mathbf{%
\check{e}}_{\alpha }(a)-u^{\beta }\ \widehat{\mathbf{\Gamma }}\mathbf{%
_{\alpha \beta }^{\gamma }\ }^{z}\mathbf{\check{e}}_{\alpha }(a)\right)
\otimes (\mathbf{\check{e}}^{\alpha }\wedge \lambda )+a\otimes d\lambda ,
\label{cdcop}
\end{equation}%
on $\mathcal{\check{W}}\otimes \Lambda ,$ where $^{z}\mathbf{\check{e}}%
_{\alpha }$ is $\mathbf{\check{e}}_{\alpha }$ redefined in $z$--variables.
This operator $\widehat{\mathbf{D}}$ is a N--adapted $\deg _{a}$--graded
derivation of the distinguished algebra $\left( \mathcal{\check{W}}\otimes
\mathbf{\Lambda ,\circ }\right) ,$ called d--algebra. Such a property
follows from (\ref{fpr}) and (\ref{cdcop})).

\begin{definition}
The Fedosov distinguished operators (d--operators) $\check{\delta}$ and $%
\check{\delta}^{-1}$ on$\ \ \mathcal{\check{W}}\otimes \mathbf{\Lambda ,}$
are defined%
\begin{equation}
\check{\delta}(a)=\ \mathbf{\check{e}}^{\alpha }\wedge \mathbf{\ }^{z}%
\mathbf{\check{e}}_{\alpha }(a),\ \mbox{and\ } \check{\delta}%
^{-1}(a)=\left\{
\begin{array}{c}
\frac{i}{p+q}z^{\alpha }\ \mathbf{\check{e}}_{\alpha }(a),\mbox{ if }p+q>0,
\\
{\qquad 0},\mbox{ if }p=q=0,%
\end{array}%
\right.  \label{feddop}
\end{equation}%
where any $a\in \mathcal{\check{W}}\otimes \mathbf{\Lambda }$ is homogeneous
w.r.t. the grading $\deg _{s}$ and $\deg _{a}$ with $\deg _{s}(a)=p$ and $%
\deg _{a}(a)=q.$
\end{definition}

The d--operators (\ref{feddop}) define the formula $a=(\check{\delta}\
\check{\delta}^{-1}+\check{\delta}^{-1}\ \check{\delta}+\sigma )(a),$ where $%
a\longmapsto \sigma (a)$ is the projection on the $(\deg _{s},\deg _{a})$%
--bihomogeneous part of $a$ of degree zero, $\deg _{s}(a)=\deg _{a}(a)=0;$ $%
\check{\delta}$ is also a $\deg _{a}$--graded derivation of the d--algebra $%
\left( \mathcal{\check{W}}\otimes \mathbf{\Lambda ,\circ }\right) .$ In
order to emphasize the almost K\"{a}hler structure, we used the canonical
almost symplectic geometric objects defined by a fixed $\mathcal{L}.$
Nevertheless, we can always change the ''Lagrangian mechanics variables''
and redefine $\mathbf{\check{\theta},}$ $\mathbf{\check{e}}_{\alpha }$ and $%
\widehat{\mathbf{\Gamma }}\mathbf{_{\alpha \beta }^{\gamma }}$ with respect
to arbitrary frame and co--frame bases using vielbeins $\mathbf{e}_{\
\underline{\alpha }}^{\alpha }$ and their duals $\mathbf{e}_{\alpha \ }^{\
\underline{\alpha }},$ defined by $e_{\ \underline{i}}^{i},e_{\ \underline{i}%
}^{a}$ and $e_{\ \underline{a}}^{a}$ satisfying (\ref{aleq}).

We can provide a "N--adapted" proof \cite{karabeg1,vqgr2} of

\begin{proposition}
\label{prthprfo}The torsion and curvature canonical d--operators of the
extension of $\widehat{\mathbf{D}}$ to $\mathcal{\check{W}}\otimes \mathbf{%
\Lambda ,}$ are computed
\begin{equation}
^{z}\widehat{\mathcal{T}}\ \doteqdot \frac{z^{\gamma }}{2}\ \check{\theta}%
_{\gamma \tau }\ \widehat{\mathbf{T}}_{\alpha \beta }^{\tau }(u)\ \mathbf{%
\check{e}}^{\alpha }\wedge \mathbf{\check{e}}^{\beta },  \label{at1}
\end{equation}%
and%
\begin{equation}
\ ^{z}\widehat{\mathcal{R}}\doteqdot \frac{z^{\gamma }z^{\varphi }}{4}\
\check{\theta}_{\gamma \tau }\ \widehat{\mathbf{R}}_{\ \varphi \alpha \beta
}^{\tau }(u)\ \mathbf{\check{e}}^{\alpha }\wedge \mathbf{\check{e}}^{\beta },
\label{ac1}
\end{equation}%
where the nontrivial coefficients of $\ \widehat{\mathbf{T}}_{\alpha \beta
}^{\tau }$ and $\ \widehat{\mathbf{R}}_{\ \varphi \alpha \beta }^{\tau }$
are defined respectively by formulas (\ref{cdtors}) and (\ref{cdcurv}).
\end{proposition}

By straightforward verifications, it follows the proof of

\begin{theorem}
\label{thprfo}The properties
\begin{equation}
\left[ \widehat{\mathbf{D}},\check{\delta}\right] =\frac{i}{v}ad_{Wick}(^{z}%
\widehat{\mathcal{T}})\mbox{ and }\ \widehat{\mathbf{D}}^{2}=-\frac{i}{v}%
ad_{Wick}(\ ^{z}\widehat{\mathcal{R}}),  \label{ffedop}
\end{equation}%
hold for the above operators, where $[\cdot ,\cdot ]$ is the $\deg _{a}$%
--graded commutator of endomorphisms of $\mathcal{\check{W}}\otimes \mathbf{%
\Lambda }$ and $ad_{Wick}$ is defined via the $\deg _{a}$--graded commutator
in $\left( \mathcal{\check{W}}\otimes \mathbf{\Lambda ,\circ }\right) .$
\end{theorem}

The formulas (\ref{ffedop}) can be redefined for any linear connection
structure on $\mathbf{V}^{2n}.$ For example, we consider how similar
formulas can be provided for the Levi--Civita connection.

\subsection{Fedosov's d--operators and the Levi--Civita connection}

For any metric structure $\mathbf{g}$ on a manifold $\mathbf{V}^{2n}\mathbf{,%
}$ the Levi--Civita connection $\bigtriangledown =\{\ _{\shortmid }\Gamma
_{\beta \gamma }^{\alpha }\}$ is by definition the unique linear connection
that is metric compatible $(\bigtriangledown g=0)$ and torsionless $( \
_{\shortmid }\mathcal{T}=0 )$. It is not a d--connection because it does not
preserve the N--connection splitting under parallel transports (\ref{whitney}%
). Let us parametrize its coefficients in the form
\begin{eqnarray*}
_{\shortmid }\Gamma _{\beta \gamma }^{\alpha } &=&\left( _{\shortmid
}L_{jk}^{i},_{\shortmid }L_{jk}^{a},_{\shortmid }L_{bk}^{i},\ _{\shortmid
}L_{bk}^{a},_{\shortmid }C_{jb}^{i},_{\shortmid }C_{jb}^{a},_{\shortmid
}C_{bc}^{i},_{\shortmid }C_{bc}^{a}\right) ,\mbox{\ where} \\
\bigtriangledown _{\mathbf{\check{e}}_{k}}(\mathbf{\check{e}}_{j}) &=&\
_{\shortmid }L_{jk}^{i}\mathbf{\check{e}}_{i}+\ _{\shortmid
}L_{jk}^{a}e_{a},\ \bigtriangledown _{\mathbf{\check{e}}_{k}}(e_{b})=\
_{\shortmid }L_{bk}^{i}\mathbf{\check{e}}_{i}+\ _{\shortmid }L_{bk}^{a}e_{a},
\\
\bigtriangledown _{e_{b}}(\mathbf{\check{e}}_{j}) &=&\ _{\shortmid
}C_{jb}^{i}\mathbf{\check{e}}_{i}+\ _{\shortmid }C_{jb}^{a}e_{a},\
\bigtriangledown _{e_{c}}(e_{b})=\ _{\shortmid }C_{bc}^{i}\mathbf{\check{e}}%
_{i}+\ _{\shortmid }C_{bc}^{a}e_{a}.
\end{eqnarray*}%
A straightforward calculation shows that the coefficients of the
Levi--Civita connection can be expressed as
\begin{eqnarray}
\ _{\shortmid }L_{jk}^{a} &=&-\widehat{C}_{jb}^{i}\check{g}_{ik}\check{g}%
^{ab}-\frac{1}{2}\check{\Omega}_{jk}^{a},\ \ _{\shortmid }L_{bk}^{i}=\frac{1%
}{2}\check{\Omega}_{jk}^{c}\check{g}_{cb}\check{g}^{ji}-\Xi _{jk}^{ih}%
\widehat{C}_{hb}^{j},  \label{lccon} \\
\ _{\shortmid }L_{jk}^{i} &=&\widehat{L}_{jk}^{i},\ _{\shortmid }L_{bk}^{a}=%
\widehat{L}_{bk}^{a}+~^{+}\Xi _{cd}^{ab}\ ^{\circ }L_{bk}^{c},\ \
_{\shortmid }C_{kb}^{i}=\widehat{C}_{kb}^{i}+\frac{1}{2}\check{\Omega}%
_{jk}^{a}\check{g}_{cb}\check{g}^{ji}+\Xi _{jk}^{ih}\widehat{C}_{hb}^{j},
\notag \\
\ _{\shortmid }C_{jb}^{a} &=&-~^{+}\Xi _{cb}^{ad}\ ^{\circ }L_{dj}^{c},\
_{\shortmid }C_{bc}^{a}=\widehat{C}_{bc}^{a},\ _{\shortmid }C_{ab}^{i}=-%
\frac{\check{g}^{ij}}{2}\left\{ \ ^{\circ }L_{aj}^{c}\check{g}_{cb}+\
^{\circ }L_{bj}^{c}\check{g}_{ca}\right\} ,  \notag
\end{eqnarray}%
where $e_{b}=\partial /\partial y^{a},$ $\check{\Omega}_{jk}^{a}$ are
computed as in (\ref{nccurv}) but for the canonical N--connection $\mathbf{%
\check{N}}$ (\ref{ncel}), $\Xi _{jk}^{ih}=\frac{1}{2}(\delta _{j}^{i}\delta
_{k}^{h}-\check{g}_{jk}\check{g}^{ih}),~^{\pm }\Xi _{cd}^{ab}=\frac{1}{2}%
(\delta _{c}^{a}\delta _{d}^{b}\pm \check{g}_{cd}\check{g}^{ab}),\ \ ^{\circ
}L_{aj}^{c}=\widehat{L}_{aj}^{c}-e_{a}(\check{N}_{j}^{c}), $ $\check{g}_{ik}$
and $\check{g}^{ab}$ are defined for the representation of the metric in
Lagrange--Finsler variables (\ref{hvmetr1}) and the normal d--connection $%
\widehat{\mathbf{\Gamma }}_{\beta \gamma }^{\alpha }=(\widehat{L}_{jk}^{i},\
^{v}\widehat{C}_{bc}^{a})$ (\ref{ndc}) is given by coefficients (\ref{cdcc}).

Let introduce the distortion d--tensor $\ _{\shortmid }Z_{\ \alpha \beta
}^{\gamma }$ with N--adapted coefficients
\begin{eqnarray}
\ _{\shortmid }Z_{jk}^{a} &=&-\widehat{C}_{jb}^{i}\check{g}_{ik}\check{g}%
^{ab}-\frac{1}{2}\check{\Omega}_{jk}^{a},~_{\shortmid }Z_{bk}^{i}=\frac{1}{2}%
\check{\Omega}_{jk}^{c}\check{g}_{cb}\check{g}^{ji}-\Xi _{jk}^{ih}~\widehat{C%
}_{hb}^{j},  \notag \\
\ _{\shortmid }Z_{jk}^{i} &=&0,\ _{\shortmid }Z_{bk}^{a}=~^{+}\Xi
_{cd}^{ab}~~^{\circ }L_{bk}^{c},_{\shortmid }Z_{kb}^{i}=\frac{1}{2}\check{%
\Omega}_{jk}^{a}\check{g}_{cb}\check{g}^{ji}+\Xi _{jk}^{ih}~\widehat{C}%
_{hb}^{j},  \label{cdeftc} \\
\ _{\shortmid }Z_{jb}^{a} &=&-~^{-}\Xi _{cb}^{ad}~~^{\circ }L_{dj}^{c},\
_{\shortmid }Z_{bc}^{a}=0,_{\shortmid }Z_{ab}^{i}=-\frac{g^{ij}}{2}\left[
~^{\circ }L_{aj}^{c}\check{g}_{cb}+~^{\circ }L_{bj}^{c}\check{g}_{ca}\right]
,  \notag
\end{eqnarray}

The next result follows from the above arguments.

\begin{proposition}
The N--adapted coefficients, of the normal d--connection and of the
distortion d--tensors define the Levi--Civita connection as
\begin{equation}
\ _{\shortmid }\Gamma _{\ \alpha \beta }^{\gamma }=\widehat{\mathbf{\Gamma }}%
_{\ \alpha \beta }^{\gamma }+\ _{\shortmid }Z_{\ \alpha \beta }^{\gamma },
\label{cdeft}
\end{equation}%
where $\ _{\shortmid }Z_{\ \alpha \beta }^{\gamma }$ are given by formulas (%
\ref{cdeft}) and h-- and v--components of $\widehat{\mathbf{\Gamma }}_{\beta
\gamma }^{\alpha }$ are given by (\ref{cdcc}).
\end{proposition}

We emphasize that all components of $\ _{\shortmid }\Gamma _{\ \alpha
\beta}^{\gamma }, \widehat{\mathbf{\Gamma }}_{\ \alpha \beta }^{\gamma }$
and $\ _{\shortmid }Z_{\ \alpha \beta }^{\gamma }$ are uniquely defined by
the coefficients of d--metric (\ref{m1}), or (equivalently) by (\ref{hvmetr1}%
) and (\ref{ncel}). The constructions can be obtained for any $n+n$
splitting on $V^{2n},$ which for suitable $\mathcal{L},$ or $\mathcal{F},$
admit a Lagrange, or Finsler, like representation of geometric objects.

By proposition \ref{prthprfo}, the expressions for  the curvature and
torsion of canonical d--operators of the extension of $\bigtriangledown $ to
$\mathcal{\check{W}}\otimes \mathbf{\Lambda ,}$ are
\begin{eqnarray}
\ _{\shortmid }^{z}\mathcal{R} &\doteqdot &\frac{z^{\gamma }z^{\varphi }}{4}%
\ \check{\theta}_{\gamma \tau }\ \ _{\shortmid }R_{\ \varphi \alpha \beta
}^{\tau }(u)\ \mathbf{\check{e}}^{\alpha }\wedge \mathbf{\check{e}}^{\beta },
\label{ac1cl} \\
\ _{\shortmid }^{z}\mathcal{T}\ &\doteqdot &\frac{z^{\gamma }}{2}\ \check{%
\theta}_{\gamma \tau }\ \ _{\shortmid }T_{\alpha \beta }^{\tau }(u)\ \mathbf{%
\check{e}}^{\alpha }\wedge \mathbf{\check{e}}^{\beta }\equiv 0,  \notag
\end{eqnarray}%
where $\ _{\shortmid }T_{\alpha \beta }^{\tau }$ $=0,$ by definition, and$\
\ _{\shortmid }R_{\ \varphi \alpha \beta }^{\tau }$ is computed with respect
to the N--adapted Lagange--Finsler canonical bases by introducing $\widehat{%
\mathbf{\Gamma }}_{\ \alpha \beta }^{\gamma }=-\ _{\shortmid }\Gamma _{\
\alpha \beta }^{\gamma }+\ _{\shortmid }Z_{\ \alpha \beta }^{\gamma },$ see (%
\ref{cdeft}), into (\ref{cdcurv}). To the N--adapted d--operator (\ref{cdcop}%
), we can associate
\begin{equation}
\widehat{\bigtriangledown }\left( a\otimes \lambda \right) \doteqdot \left(
\mathbf{\check{e}}_{\alpha }(a)-u^{\beta }\ _{\shortmid }\Gamma \mathbf{%
_{\alpha \beta }^{\gamma }\ }^{z}\mathbf{\check{e}}_{\alpha }(a)\right)
\otimes (\mathbf{\check{e}}^{\alpha }\wedge \lambda )+a\otimes d\lambda ,
\label{lcexop}
\end{equation}%
on $\mathcal{\check{W}}\otimes \Lambda ,$ where $^{z}\mathbf{\check{e}}%
_{\alpha }$ is $\mathbf{\check{e}}_{\alpha }$ redefined in $z$--variables.
This almost symplectic connection $\widehat{\bigtriangledown }$ is
torsionles and, in general, is not adapted to the N--connection structures.

\begin{corollary}
For the Levi--Civita connection $\bigtriangledown =\{\ _{\shortmid }\Gamma
_{\beta \gamma }^{\alpha }\}$ on a N--anholo\-no\-mic manifold $\mathbf{V}%
^{2n},$ we have:
\begin{equation*}
\left[ \widehat{\bigtriangledown },\check{\delta}\right] =0\mbox{ and }\
\widehat{\bigtriangledown }^{2}=-\frac{i}{v}ad_{Wick}(\ _{\shortmid }^{z}%
\mathcal{R}),
\end{equation*}%
where $\widehat{\bigtriangledown }$ is defined by formula (\ref{lcexop}), $\
_{\shortmid }^{z}\mathcal{R}$ is given by (\ref{ac1cl}), $[\cdot ,\cdot ]$
is the $\deg _{a}$--graded commutator of endomorphisms of $\mathcal{\check{W}%
}\otimes \mathbf{\Lambda }$ and $ad_{Wick}$ is defined via the $\deg _{a}$%
--graded commutator in $\left( \mathcal{\check{W}}\otimes \mathbf{\Lambda
,\circ }\right) .$
\end{corollary}

\begin{proof}
It is a straightforward consequence of the Theorem \ref{thprfo} for the
Levi--Civita and curvature operators extended on $\mathcal{\check{W}}\otimes
\Lambda .$ $\square $
\end{proof}

\vskip5pt

Prescribing a $n+n$ splitting on $\mathbf{V}^{2n}$, we can work equivalently
with any metric compatible linear connection structure which is N--adapted,
or not, if such a connection is completely defined by the (pseudo)
Riemannian metric structure. It is preferable to use the approach with the
normal d--connection because this way we have both an almost symplectic
analogy and Lagrange, or Finsler, like interpretation of geometric objects.
In standard classical gravity, in order to solve some physical problems, it
is more convenient to work with the Levi--Civita connection or its spin like
representations (for instance, in the Einstein--Dirac theory). The
self--dual and further generalizations to Ashtekar variables are more
convenient, respectively, in canonical ADN classical and quantum gravity
and/or loop quantum gravity.

It should be noted that the formulas for Fedosov's d--operators and their
properties do not depend in explicit form on generating functions $\mathcal{L%
},$ or $\mathcal{F}.$ Such a function may be formally introduced for
elaborating a Lagrange mechanics, or Finsler, modeling for a (pseudo)
Riemannian space with a general $n+n$ nonholonomic splitting. This way, we
emphasize that the Fedosov's approach is valid for various type of (pseudo)
Riemann, Riemann--Cartan, Lagrange--Finsler, almost K\"{a}hler and other
types of holonomic and nonholonic manifolds used for geometrization of
mechanical and field models. Nevertheless, the constructions are performed
in a general form and the final results do not depend on any ''background''
structures. We conclude that $3+1$ fibration approaches are more natural for
loop quantum gravity, but the models with nonholonomic $2+2$ splitting
result in almost K\"{a}hler quantum models; althought both types of
quantization, loop and deformation, provide background independent
constructions.

\section{Deformation Quantization of Einstein and Lagrange Spaces}

Formulating a (pseudo) Riemannian geometry in Lagrange--Finsler variables,
we can quantize the metric, frame and linear connection structures following
standard methods for deformation quantization of almost K\"{a}hler
manifolds. The goal of this section is to provide the main Fedosov type
results for such constructions and to show how the Einstein manifolds can be
encoded into the topological structure of such quantized nonholonomic spaces.

\subsection{Fedosov's theorems for normal d--connections}

The third main result of this work will be stated below by three theorems
for the normal d--connection (equivalently, canonical almost symplectic
structure) $\widehat{\mathbf{D}}\equiv \ _{\theta }\widehat{\mathbf{D}}$ (%
\ref{ndc}). Such results were formulated originally in Fedosov's works \cite%
{fedos1,fedos2} and generalized, for instance, for various classes of metric
compatible affine connections, canonical Lagrange--Finsler connections and
effective locally anisotropic quantum gravities, see Refs. \cite%
{karabeg1,vqgr1,vqgr2,vqgr3}.

\begin{theorem}
\label{th3a}Any (pseudo) Riemanian metric $\mathbf{g}$ (\ref{m1})
(equivalently, $\mathbf{g=\check{g}}$ (\ref{hvmetr1})) defines a flat normal
Fedosov d--connec\-ti\-on
\begin{equation*}
\ \widehat{\mathcal{D}}\doteqdot -\ \check{\delta}+\widehat{\mathbf{D}}-%
\frac{i}{v}ad_{Wick}(r)
\end{equation*}%
satisfying the condition $\widehat{\mathcal{D}}^{2}=0,$ where the unique
element $r\in $ $\mathcal{\check{W}}\otimes \mathbf{\Lambda ,}$ $\deg
_{a}(r)=1,$ $\check{\delta}^{-1}r=0,$ solves the equation
\begin{equation*}
\ \check{\delta}r=\widehat{\mathcal{T}}\ +\widehat{\mathcal{R}}+\widehat{%
\mathbf{D}}r-\frac{i}{v}r\circ r
\end{equation*}%
and this element can be computed recursively with respect to the total
degree $Deg$ as follows:%
\begin{eqnarray*}
r^{(0)} &=&r^{(1)}=0, r^{(2)}=\check{\delta}^{-1}\widehat{\mathcal{T}},
r^{(3)}=\ \check{\delta}^{-1}\left( \widehat{\mathcal{R}}+\widehat{\mathbf{D}%
}r^{(2)}-\frac{i}{v}r^{(2)}\circ r^{(2)}\right) , \\
r^{(k+3)} &=&\ \ \check{\delta}^{-1}\left( \widehat{\mathbf{D}}r^{(k+2)}-%
\frac{i}{v}\sum\limits_{l=0}^{k}r^{(l+2)}\circ r^{(l+2)}\right) ,k\geq 1,
\end{eqnarray*}%
where by $a^{(k)}$ we denoted the $Deg$--homogeneous component of degree $k$
of an element $a\in $ $\ \mathcal{\check{W}}\otimes \mathbf{\Lambda }.$
\end{theorem}

\begin{proof}
It follows from straightforward verifications of the property $\widehat{%
\mathcal{D}}^{2}=0$ using for $r$ formal series of type (\ref{formser}) and
the formulas for N--adapted coefficients: (\ref{cdcc}) for $\widehat{\mathbf{%
D}},$ (\ref{cdtors}) for $\widehat{\mathcal{T}},$ (\ref{cdcurv}) for $%
\widehat{\mathcal{R}},$ and the properties of Fedosov's d--operators (\ref%
{feddop}) stated by Theorem \ref{thprfo}. The length of this paper does not
allow us to present such a tedious calculation which is a N--adapted version
for corresponding ''hat'' operators, see the related work in \cite%
{fedos1,fedos2,karabeg1}.$\square $
\end{proof}

\vskip5pt

The procedure of deformation quantization is related to the definition of a
star--product which in our approach can be defined canonically as in \cite%
{karabeg1} because the normal d--connection $\widehat{\mathbf{D}}$ is a
N--adapted variant of the affine and almost symplectic connection considered
in that work. This provides a proof for

\begin{theorem}
\label{th3b}A star--product on the almost K\"{a}hler model of a (pseudo)
Riemannian space in Lagrange--Finsler variables is defined on $C^{\infty }(%
\mathbf{V}^{2n})[[v]]$ by formula
\begin{equation*}
\ ^{1}f\ast \ ^{2}f\doteqdot \sigma (\tau (\ ^{1}f))\circ \sigma (\tau (\
^{2}f)),
\end{equation*}%
where the projection $\sigma :\mathcal{\check{W}}_{\widehat{\mathcal{D}}%
}\rightarrow C^{\infty }(\mathbf{V}^{2n})[[v]]$ onto the part of $\deg _{s}$%
--degree zero is a bijection and the inverse map $\tau :C^{\infty }(\mathbf{V%
}^{2n})[[v]]\rightarrow \mathcal{\check{W}}_{\widehat{\mathcal{D}}}$ can be
calculated recursively w.r..t the total degree $Deg,$%
\begin{eqnarray*}
\tau (f)^{(0)} &=&f\mbox{\ and, for \ }k\geq 0, \\
\tau (f)^{(k+1)} &=&\ \check{\delta}^{-1}\left( \widehat{\mathbf{D}}\tau
(f)^{(k)}-\frac{i}{v}\sum\limits_{l=0}^{k}ad_{Wick}(r^{(l+2)})(\tau
(f)^{(k-l)})\right) .
\end{eqnarray*}
\end{theorem}

We denote by $\ ^{f}\xi $ the Hamiltonian vector field corresponding to a
function $f\in C^{\infty }(\mathbf{V}^{2n})$ on space $(\mathbf{V}^{2n},%
\check{\theta})$ and consider the antisymmetric part $\ ^{-}C(\ ^{1}f,\
^{2}f)\ \doteqdot \frac{1}{2}\left( C(\ ^{1}f,\ ^{2}f)-C(\ ^{2}f,\
^{1}f)\right) $ of bilinear operator $C(\ ^{1}f,\ ^{2}f).$ We say that a
star--product (\ref{starp}) is normalized if $\ _{1}C(\ ^{1}f,\ ^{2}f)=\frac{%
i}{2}\{\ ^{1}f,\ ^{2}f\},$ where $\{\cdot ,\cdot \}$ is the Poisson bracket.
For the normalized $\ast ,$ the bilinear operator $\ _{2}^{-}C$ defines a de
Rham--Chevalley 2--cocycle, when there is a unique closed 2--form $\ \check{%
\varkappa}$ such that%
\begin{equation}
\ _{2}C(\ ^{1}f,\ ^{2}f)=\frac{1}{2}\ \check{\varkappa}(\ ^{f_{1}}\xi ,\
^{f_{2}}\xi )  \label{c2}
\end{equation}%
for all $\ ^{1}f,\ ^{2}f\in C^{\infty }(\mathbf{V}^{2n}).$ This is used to
introduce $c_{0}(\ast )\doteqdot \lbrack \check{\varkappa}]$ as the
equivalence class.

A straightforward computation of $\ _{2}C$ from (\ref{c2}) and the results
of Theorem \ref{th3b} provide the proof of

\begin{lemma}
\label{lem1}The unique 2--form defined by the normal d--connection can be
computed as
\begin{equation*}
\check{\varkappa}=-\frac{i}{8}\mathbf{\check{J}}_{\tau }^{\ \alpha ^{\prime
}}\widehat{\mathcal{R}}_{\ \alpha ^{\prime }}^{\tau }-\frac{i}{6}d\left(
\mathbf{\check{J}}_{\tau }^{\ \alpha ^{\prime }}\widehat{\mathbf{T}}_{\
\alpha ^{\prime }\beta }^{\tau }\ \mathbf{\check{e}}^{\beta }\right) ,
\end{equation*}%
where the coefficients of the curvature and torsion 2--forms of the normal
d--connection 1--form are given respectively by formulas (\ref{cform}) and (%
\ref{tform}).
\end{lemma}

We now define another canonical class $\check{\varepsilon},$ for $\ ^{%
\check{N}}T\mathbf{V}^{2n}=h\mathbf{V}^{2n}\oplus v\mathbf{V}^{2n},$ where
the left label indicates that the tangent bundle is split nonholonomically
by the canonical N--connection structure $\mathbf{\check{N}}.$ We can
perform a distinguished complexification of such second order tangent
bundles in the form $T_{\mathbb{C}}\left( \ ^{\check{N}}T\mathbf{V}%
^{2n}\right) =T_{\mathbb{C}}\left( h\mathbf{V}^{2n}\right) \oplus T_{\mathbb{%
C}}\left( v\mathbf{V}^{2n}\right) $ and introduce $\ \check{\varepsilon}$ as
the first Chern class of the distributions $T_{\mathbb{C}}^{\prime }\left( \
^{N}T\mathbf{V}^{2n}\right) =T_{\mathbb{C}}^{\prime }\left( h\mathbf{V}%
^{2n}\right) \oplus T_{\mathbb{C}}^{\prime }\left( v\mathbf{V}^{2n}\right) $
of couples of vectors of type $(1,0)$ both for the h-- and v--parts. In
explicit form, we can calculate $\check{\varepsilon}$ by using the
d--connection $\widehat{\mathbf{D}}$ and the h- and v--projections $h\Pi =%
\frac{1}{2}(Id_{h}-iJ_{h})$ and $v\Pi =\frac{1}{2}(Id_{v}-iJ_{v}),$ where $%
Id_{h}$ and $Id_{v}$ are respective identity operators and $J_{h}$ and $%
J_{v} $ are almost complex operators, which are projection operators onto
corresponding $(1,0)$--subspaces. Introducing the matrix $\left( h\Pi ,v\Pi
\right) \ \widehat{\mathcal{R}}\left( h\Pi ,v\Pi \right) ^{T},$ where $%
(...)^{T}$ means transposition, as the curvature matrix of the N--adapted
restriction of $\ $of the normal d--connection$\ \widehat{\mathbf{D}}$ to $%
T_{\mathbb{C}}^{\prime }\left( \ ^{\check{N}}T\mathbf{V}^{2n}\right) ,$ we
compute the closed Chern--Weyl form
\begin{equation}
\check{\gamma}=-iTr\left[ \left( h\Pi ,v\Pi \right) \widehat{\mathcal{R}}%
\left( h\Pi ,v\Pi \right) ^{T}\right] =-iTr\left[ \left( h\Pi ,v\Pi \right)
\widehat{\mathcal{R}}\right] =-\frac{1}{4}\mathbf{\check{J}}_{\tau }^{\
\alpha ^{\prime }}\widehat{\mathcal{R}}_{\ \alpha ^{\prime }}^{\tau }.
\label{aux4}
\end{equation}%
We get that the canonical class is $\check{\varepsilon}\doteqdot \lbrack
\check{\gamma}],$ which proves the

\begin{theorem}
\label{th3c}The zero--degree cohomology coefficient $c_{0}(\ast )$ for the
almost K\"{a}hler model of a (pseudo) Riemannian space defined by d--tensor $%
\mathbf{g}$ (\ref{m1}) (equivalently, by $\mathbf{\check{g}}$ (\ref{hvmetr1}%
)) is computed $c_{0}(\ast )=-(1/2i)\ \check{\varepsilon}.$
\end{theorem}

The coefficient $c_{0}(\ast )$ can be similarly computed for the case when a
metric of type (\ref{m1}) is a solution of the Einstein equations and this
zero--degree coefficient defines certain quantum properties of the
gravitational field. A more rich geometric structure should be considered if
we define a value similar to $c_{0}(\ast )$ encoding the information about
Einstein manifolds deformed into corresponding quantum configurations.

\subsection{The zero--degree cohomology coefficient for Einstein ma\-nifolds}

\label{ssensp}The priority of deformation quantization is that we can
elaborate quantization schemes when metric, vielbein and connection fields
are not obligatory subjected to satisfy certain field equations and/or
derived by a variational procedure. For instance, such geometric and/or BRST
quantization approaches were proposed in Ref. \cite{lyakh1,lyakh2}. On the
other hand, in certain canonical and loop quantization models, the
gravitational field equations are considered as the starting point for
deriving a quantization formalism. In such cases, the Einstein equations are
expressed into ''lapse'' and ''shift'' (and/or generalized Ashtekar)
variables and the quantum variant of the gravitational field equations is
prescribed to be in the form of Wheeler De Witt equations (or corresponding
systems of constraints in complex/real generalized connection and dreibein
variables). In this section, we analyze the problem of encoding the Einstein
equations into a geometric formalism of nonholonomic deformation
quantization.

\subsubsection{Gravitational field equations}

For any d--connection $\mathbf{D=\{\Gamma \},}$ we can define the Ricci
tensor $Ric(\mathbf{D})=\{\mathbf{R}_{\ \beta \gamma }\doteqdot \mathbf{R}%
_{\ \beta \gamma \alpha }^{\alpha }\}$ and the scalar curvature $\
^{s}R\doteqdot \mathbf{g}^{\alpha \beta }\mathbf{R}_{\alpha \beta }$ ($%
\mathbf{g}^{\alpha \beta }$ being the inverse matrix to $\mathbf{g}_{\alpha
\beta }$ (\ref{m1})). If a d--connection is uniquely determined by a metric
in a unique metric compatible form, $\mathbf{Dg}=0,$ (in general, the
torsion of $\mathbf{D}$ is not zero, but induced canonically by the
coefficients of $\mathbf{g),}$ we can postulate in straightforward form the
field equations
\begin{equation}
\mathbf{R}_{\ \beta }^{\underline{\alpha }}-\frac{1}{2}(\ ^{s}R+\lambda )%
\mathbf{e}_{\ \beta }^{\underline{\alpha }}=8\pi G\mathbf{T}_{\ \beta }^{%
\underline{\alpha }},  \label{deinsteq}
\end{equation}%
where $\mathbf{T}_{\ \beta }^{\underline{\alpha }}$ is the effective
energy--momentum tensor, $\lambda $ is the cosmological constant, $G$ is the
Newton constant in the units when the light velocity $c=1,$ and $\mathbf{e}%
_{\ \beta }=\mathbf{e}_{\ \beta }^{\underline{\alpha }}\partial /\partial u^{%
\underline{\alpha }}$ is the N--elongated operator (\ref{dder}).

Let us consider the absolute antisymmetric tensor $\epsilon _{\alpha \beta
\gamma \delta }$ and effective source 3--form
\begin{equation*}
\mathcal{T}_{\ \beta }=\mathbf{T}_{\ \beta }^{\underline{\alpha }}\ \epsilon
_{\underline{\alpha }\underline{\beta }\underline{\gamma }\underline{\delta }%
}du^{\underline{\beta }}\wedge du^{\underline{\gamma }}\wedge du^{\underline{%
\delta }}
\end{equation*}%
and express the curvature tensor $\mathcal{R}_{\ \gamma }^{\tau }=\mathbf{R}%
_{\ \gamma \alpha \beta }^{\tau }\ \mathbf{e}^{\alpha }\wedge \ \mathbf{e}%
^{\beta }$ of $\mathbf{\Gamma }_{\ \beta \gamma }^{\alpha }=\ _{\shortmid
}\Gamma _{\ \beta \gamma }^{\alpha }-\ Z_{\ \beta \gamma }^{\alpha }$ as $%
\mathcal{R}_{\ \gamma }^{\tau }=\ _{\shortmid }\mathcal{R}_{\ \gamma }^{\tau
}-\mathcal{Z}_{\ \gamma }^{\tau },$ where $\ _{\shortmid }\mathcal{R}_{\
\gamma }^{\tau }$ $=\ _{\shortmid }R_{\ \gamma \alpha \beta }^{\tau }\
\mathbf{e}^{\alpha }\wedge \ \mathbf{e}^{\beta }$ is the curvature 2--form
of the Levi--Civita connection $\nabla $ and the distortion of curvature
2--form $\mathcal{Z}_{\ \gamma }^{\tau }$ is defined by $\ Z_{\ \beta \gamma
}^{\alpha }.$ For the gravitational $\left( \mathbf{e,\Gamma }\right) $ and
matter $\mathbf{\phi }$ fields, we consider the effective action
\begin{equation*}
S[\mathbf{e,\Gamma ,\phi }]=\ ^{gr}S[\mathbf{e,\Gamma }]+\ ^{matter}S[%
\mathbf{e,\Gamma ,\phi }].
\end{equation*}

\begin{theorem}
\label{theq}The equations (\ref{deinsteq}) can be represented as 3--form
equations%
\begin{equation}
\epsilon _{\alpha \beta \gamma \tau }\left( \mathbf{e}^{\alpha }\wedge
\mathcal{R}^{\beta \gamma }+\lambda \mathbf{e}^{\alpha }\wedge \ \mathbf{e}%
^{\beta }\wedge \ \mathbf{e}^{\gamma }\right) =8\pi G\mathcal{T}_{\ \tau }
\label{einsteq}
\end{equation}%
following from the action by varying the components of $\mathbf{e}_{\ \beta
},$ when%
\begin{eqnarray*}
\mathcal{T}_{\ \tau }&=&\ ^{m}\mathcal{T}_{\ \tau }+\ ^{Z}\mathcal{T}_{\
\tau }, \\
\ ^{m}\mathcal{T}_{\ \tau } &=&\ ^{m}\mathbf{T}_{\ \tau }^{\underline{\alpha
}}\epsilon _{\underline{\alpha }\underline{\beta }\underline{\gamma }%
\underline{\delta }}du^{\underline{\beta }}\wedge du^{\underline{\gamma }%
}\wedge du^{\underline{\delta }}, \\
\ ^{Z}\mathcal{T}_{\ \tau } &=&\left( 8\pi G\right) ^{-1}\mathcal{Z}_{\ \tau
}^{\underline{\alpha }}\epsilon _{\underline{\alpha }\underline{\beta }%
\underline{\gamma }\underline{\delta }}du^{\underline{\beta }}\wedge du^{%
\underline{\gamma }}\wedge du^{\underline{\delta }},
\end{eqnarray*}%
where $\ ^{m}\mathbf{T}_{\ \tau }^{\underline{\alpha }}=\delta \
^{matter}S/\delta \mathbf{e}_{\underline{\alpha }}^{\ \tau }$ are equivalent
to the usual Einstein equations for the Levi--Civita connection $\nabla ,$%
\begin{equation*}
\ _{\shortmid }\mathbf{R}_{\ \beta }^{\underline{\alpha }}-\frac{1}{2}(\
_{\shortmid }^{s}R+\lambda )\mathbf{e}_{\ \beta }^{\underline{\alpha }}=8\pi
G\ ^{m}\mathbf{T}_{\ \beta }^{\underline{\alpha }}.
\end{equation*}
\end{theorem}

\begin{proof}
It is a usual textbook and/or differential form calculus (see, for instance, %
\cite{mtw,rovelli}), but with respect to N--adapted bases (\ref{dder}) and (%
\ref{ddif}) for a metric compatible d--connection (\ref{deinsteq}). $\square
$
\end{proof}

\vskip3pt

As a particular case, for the Einstein gravity in Lagrange--Finsler
variables, we obtain:

\begin{corollary}
The vacuum Einstein equations, with cosmological constant in terms of the
canonical N--adapted vierbeins and normal d--connection, are%
\begin{equation}
\epsilon _{\alpha \beta \gamma \tau }\left( \mathbf{\check{e}}^{\alpha
}\wedge \widehat{\mathcal{R}}^{\beta \gamma }+\lambda \mathbf{\check{e}}%
^{\alpha }\wedge \mathbf{\check{e}}^{\beta }\wedge \ \mathbf{\check{e}}%
^{\gamma }\right) =8\pi G\ ^{Z}\widehat{\mathcal{T}}_{\ \tau },
\label{veinst1}
\end{equation}%
or, in terms of the Levi--Civita connection%
\begin{equation*}
\epsilon _{\alpha \beta \gamma \tau }\left( \mathbf{\check{e}}^{\alpha
}\wedge \ _{\shortmid }\mathcal{R}^{\beta \gamma }+\lambda \mathbf{\check{e}}%
^{\alpha }\wedge \mathbf{\check{e}}^{\beta }\wedge \ \mathbf{\check{e}}%
^{\gamma }\right) =0.
\end{equation*}
\end{corollary}

\begin{proof}
The conditions of the mentioned Theorem \ref{theq} are redefined for the
co--frames $\mathbf{\check{e}}^{\alpha }$ elongated by the canonical
N--connection (\ref{ncel}), deformation of linear connections (\ref{cdeft})
and curvature (\ref{cdcurv}) with deformation of curvature 2--form of type
\begin{equation}
\widehat{\mathcal{R}}_{\ \gamma }^{\tau }=\ _{\shortmid }\mathcal{R}_{\
\gamma }^{\tau }-\widehat{\mathcal{Z}}_{\ \gamma }^{\tau }.  \label{def2}
\end{equation}%
We put ''hat'' on $\ ^{Z}\widehat{\mathcal{T}}_{\ \tau }$ because this value
is computed using the normal d--connection. $\square $
\end{proof}

\vskip3pt

Using formulas (\ref{veinst1}) and (\ref{def2}), we can write
\begin{equation}
\widehat{\mathcal{R}}^{\beta \gamma }=-\lambda \mathbf{\check{e}}^{\beta
}\wedge \ \mathbf{\check{e}}^{\gamma }-\widehat{\mathcal{Z}}^{\beta \gamma }%
\mbox{ and }\ _{\shortmid }\mathcal{R}^{\beta \gamma }=-\lambda \mathbf{%
\check{e}}^{\beta }\wedge \ \mathbf{\check{e}}^{\gamma }.  \label{aux3}
\end{equation}%
Such formulas are necessary for encoding the vacuum field equations into the
cohomological structure of quantum almost K\"{a}hler model of Einstein
gravity.

\subsubsection{The Chern--Weyl form and Einstein equations}

Introducing the formulas (\ref{veinst1}) and (\ref{aux3}) into the
conditions of Lemma \ref{lem1} and Theorem \ref{th3c}, we obtain the forth
main result in this work:

\begin{theorem}
\label{th4r}The zero--degree cohomology coefficient $c_{0}(\ast )$ for the
almost K\"{a}hler model of an Einstein space defined by a d--tensor $\mathbf{%
g}$ (\ref{m1}) (equivalently, by $\mathbf{\check{g}}$ (\ref{hvmetr1})) as a
solution of \ (\ref{veinst1}) is $c_{0}(\ast )=-(1/2i)\ \check{\varepsilon},$
for $\check{\varepsilon}\doteqdot \lbrack \check{\gamma}],$ where
\begin{equation}
\check{\gamma}=\frac{1}{4}\mathbf{\check{J}}_{\tau \alpha }^{\ }\left(
-\lambda \mathbf{\check{e}}^{\tau }\wedge \ \mathbf{\check{e}}^{\alpha }+%
\widehat{\mathcal{Z}}^{\tau \alpha }\right) .  \label{cwform}
\end{equation}
\end{theorem}

\begin{proof}
We sketch the key points of the proof which follows from (\ref{aux4}) and (%
\ref{aux3}). It should be noted that for $\lambda =0$ the 2--form $\widehat{%
\mathcal{Z}}^{\tau \alpha }$ is defined by the deformation d--tensor from
the Levi--Civita connection to the normal d--connection (\ref{cdeft}), see
formulas (\ref{cdeftc}). Such objects are defined by classical vacuum
solutions of the Einstein equations. We conclude that $c_{0}(\ast )$ encodes
the vacuum Einstein configurations, in general, with nontrivial constants
and their quantum deformations. $\square $
\end{proof}

\vskip5pt

If the Wheeler De Witt equations represent a quantum version of the Einstein
equations for loop quantum gravity (see discussions in Refs. \cite%
{rovelli,asht3,thiem1}), the Chern--Weyl 2--form (\ref{cwform}) can be used
to define the quantum version of Einstein equations (\ref{einsteq}) in the
deformation quantization approach:

\begin{corollary}
In Lagrange--Finsler variables, the quantum field equations corresponding to
Einstein's general relativity are
\begin{equation}
\mathbf{\check{e}}^{\alpha }\wedge \check{\gamma}=\epsilon ^{\alpha \beta
\gamma \tau }2\pi G\mathbf{\check{J}}_{\beta \gamma }\widehat{\mathcal{T}}%
_{\ \tau }\ -\frac{\lambda }{4}\mathbf{\check{J}}_{\beta \gamma }\mathbf{%
\check{e}}^{\alpha }\wedge \mathbf{\check{e}}^{\beta }\wedge \ \mathbf{%
\check{e}}^{\gamma }.  \label{aseq}
\end{equation}
\end{corollary}

\begin{proof}
Multiplying $\mathbf{\check{e}}^{\alpha }\wedge $ to (\ref{cwform}) written
in Lagrange--Finsler variables and taking into account (\ref{einsteq}),
re--written also in the form adapted to the canonical N--connection, and
introducing the almost complex operator $\mathbf{\check{J}}_{\beta \gamma },$
we get the almost symplectic form of Einstein's equations (\ref{aseq}). $%
\square $
\end{proof}

\vskip5pt

It should be noted that even in the vacuum case, when $\lambda =0,$ the
2--form $\check{\gamma}$ (\ref{cwform}) from (\ref{aseq}) is not zero but
defined by $\widehat{\mathcal{T}}_{\ \tau }=\ ^{Z}\widehat{\mathcal{T}}_{\
\tau }.$

Finally, we emphasize that an explicit computation of $\check{\gamma}$ for
nontrivial matter fields has yet to be performed for a deformation
quantization model in which interacting gravitational and matter fields are
geometrized in terms of an almost K\"{a}hler model defined for spinor and
fiber bundles on spacetime. This is a subject for further investigations.

\section{Conclusions and Discussion}

So far we have dealt with the deformation quantization of general relativity
in Lagrange--Finsler variables, inducing a nonholonomic 2+2 splitting, as a
geometric alternative to the 3+1 setting to loop quantum gravity. In our
approach, the methods of geometric mechanics and Finsler geometry are
canonically combined in order to convert general relativity into an almost K%
\"{a}hler structure for which a well defined formalism of geometric
quantization exists. The formalism is elaborated in a background independent
and nonperturbative form, for Lortentzian gravitational fields on four
dimensional manifolds, with possible extensions to extra dimensions.

There are many physical interesting formulations of gravity theories in
differential form, with tetrad, spinor and different connection variables.
For instance, the 3+1 splitting and Ashtekar variables resulted in a
similarity with Yang Mills theory and allowed us to simplify the constraints
and provide consistent loop quantum gravity formulations. In order to
construct almost K\"{a}hler models of classical and quantum gravity, it is
more convenient to use a 2+2 splitting with prescribed nonholonomic frame
structures when the metric and linear connection transform into canonical
symplectic forms and connections. The main advantage of this approach is
that we do not have to solve constraint equations and can apply directly the
methods of deformation quantization. We can work both with tetrad and
connection variables, and the methods can be generalized for Lagrange and
Finsler spaces, almost K\"{a}hler and almost Poisson structures,
nonsymmetric metrics and noncommutative geometries.

The facts that Einstein's theory is diffeomorphic invariant and preserves
local Lorentz invariance are crucial features at the classical level and
provide strong motivations to preserve such symmetries at the quantum level.
We emphasize here that by prescribing a distribution by defining a
nonholonomic frame structure with associated nonlinear connection on a
(semi) Riemannian manifold, we do not break general covariance. We chose to
work with a class of frame transforms that did not affect the general
properties of classical and quantum gravity theories. In general, all
constructions can be re--defined for arbitrary frames and coordinate systems.

Let us outline the main results (four) of this paper. The first one is
provided by Theorem \ref{mth1}, stating that any (pseudo) Riemannian space
can be described equivalently in terms of effective Lagrange (or Finsler)
variables. This allows us to prove the second main result formulated in
Theorem \ref{thmr2}: Having chosen a generating Lagrange (Finsler) function $%
\mathcal{L}(x,y)$ (or $\mathcal{F}(x,y))$ on a (pseudo) Riemannian manifold,
we can model this space as an almost K\"{a}hler geometry. Conventionally,
the third main result is split into three (Theorems \ref{th3a}, \ref{th3b}
and \ref{th3c}) Fedosov's theorems for the normal distinguished connection
in general relativity and its deformation quantization. We have introduced
the normal Fedosov's distinguished operators, constructed the star product
and computed the zero--degree cohomology coefficient for the almost K\"{a}%
hler models of (pseudo) Riemannian and Einstein spaces. The fourth result is
given by Theorem \ref{th4r}, which states that a corresponding zero--degree
cohomology coefficient also encodes the information about solutions of
Einstein equations. This allows us to introduce, in Lagrange--Finsler
variables and using a related Chern--Cartan form, the quantum field
equations corresponding to the classical gravitational equations in general
relativity.

In this paper, we only concluded the first steps \cite{vqgr1,vqgr2,vqgr3,esv}
towards a consistent deformation quantization of gravity using the nonlinear
connection formalism and the methods of Lagrange--Finsler geometry in
Einstein gravity and generalizations. Many details of more complete
constructions are still lacking. For example, we provided only the
transformations suitable to implement deformation quantization methods but
we have not discussed a de--quantization procedure and relevance to the
semiclassical limit of such gravitational models. There are many unsolved
problems pertaining canonical and quantum loop quantizations. It is very
likely that non--holonomic geometry quantization methods of gravity and its
relation to loop gravity, canonical and perturbative approaches,
noncommutative generalizations and applications to modern cosmology and
gravity physics will play an important role in future investigations.

\vskip3pt

\textbf{Acknowledgement: } The work was performed during a visit at Fields Institute. Author is grateful to Douglas Singleton and Andrea Arias de Gill from California State University at Fresno, USA, for substantial help on this paper.

\end{document}